\documentclass{aa} 

\usepackage{graphicx}
\usepackage{txfonts}
\usepackage{longtable}
\usepackage{chemformula}
\usepackage{placeins}
\usepackage{caption}
\usepackage{subcaption}
\newcommand{\kms}{\,km\,s$^{-1}$} 
\newcommand{\ms}{\,m\,s$^{-1}$} 
\newcommand{\gcm}{\,g\,cm$^{-3}$} 

\newcommand{\mearth}{$M_{\oplus}$}
\newcommand{\rearth}{$R_{\oplus}$}
\newcommand{\msun}{$M_{\odot}$}

\DeclareSymbolFont{UPM}{U}{eur}{m}{n}
\DeclareMathSymbol{\umu}{0}{UPM}{"16}
\let\oldumu=\umu
\renewcommand\umu{\ifmmode\oldumu\else$\oldumu$\fi}
\newcommand\micro{\umu}
\newcommand\micron{\micro\rm m}
\newcommand\microns{\micron}

\begin{document} 

    \title{TOI-837\,$b$: characterisation, formation and evolutionary history of an infant warm Saturn-mass planet}

   \author{M.~Damasso
          \inst{1}
          \and D.~Polychroni\inst{2}
          \and D.~Locci\inst{3}
          \and D.~Turrini\inst{1}
          \and A.~Maggio\inst{3}
          \and P.~E.~Cubillos\inst{4,1}
          \and M.~Baratella\inst{5}
          \and K.~Biazzo\inst{6}
          \and S.~Benatti\inst{3}       
          \and G.~Mantovan\inst{7}
          \and D.~Nardiello\inst{7,8}
          \and S.~Desidera\inst{8}
          \and A.~S.~Bonomo\inst{1}
          \and M.~Pinamonti\inst{1}
          \and L.~Malavolta\inst{7}
          \and F.~Marzari\inst{7,8}
          \and A.~Sozzetti\inst{1}
          \and R.~Spinelli\inst{3}
          }

   \institute{INAF -- Osservatorio Astrofisico di Torino, Via Osservatorio 20, I-10025 Pino Torinese, Italy\\
           	\email{mario.damasso@inaf.it}
            \and INAF -- Osservatorio Astronomico di Trieste, Via Giambattista
Tiepolo, 11, I-34131, Trieste (TS), Italy
            \and INAF -- Osservatorio Astronomico di Palermo, Piazza del Parlamento 1, I-90134, Palermo, Italy
            \and Space Research Institute, Austrian Academy of Sciences, Schmiedlstrasse 6, 8042, Graz, Austria
            \and ESO-European Southern Observatory, Alonso de Cordova 3107, Vitacura, Santiago, Chile
            \and INAF -- Osservatorio Astronomico di Roma, Via Frascati 33, 00078 -- Monte Porzio Catone (Roma), Italy
            \and Dipartimento di Fisica e Astronomia "G. Galilei"-- Universt\`a degli Studi di Padova, Vicolo dell'Osservatorio 3, I-35122 Padova
            \and INAF - Osservatorio Astronomico di Padova, Vicolo dell’Osservatorio 5, IT-35122, Padova, Italy
             }

   \date{}

 
  \abstract
   {The detection and characterisation of planets younger than $\sim$100 Myr offer the opportunity to get snap-shots of systems right after their formation, and where the main evolutionary processes that sculpt mature planetary systems are still ongoing. The sample of known infant exoplanets is currently scarce, and dedicated surveys are required to increase their number.}
   {We aim to determine the fundamental properties of the $\sim$35 Myr old star TOI-837 and its close-in Saturn-sized planet, and to investigate the system's formation and evolutionary history.}
   {We analysed TESS photometry and HARPS spectroscopic data, measured stellar and planetary parameters, and characterised the stellar activity. We performed population synthesis simulations to track the formation history of TOI-837\,$b$, and to reconstruct its possible internal structure. We investigated the planetary atmospheric evolution through photo-evaporation, and quantified the prospects for atmospheric characterisation with JWST. }
   {TOI-837\,$b$ has radius and mass similar to those of Saturn ($r_b$=9.71$^{+0.93}_{-0.60}$ \rearth, $m_b$=116$^{+17}_{-18}$ \mearth, and $\rho_b$=0.68$^{+0.20}_{-0.18}$ \gcm), on a primordial circular orbit. Population synthesis and early migration simulations suggest that the planet could have originated between 2--4 au, and have either a large and massive core, or a smaller Saturn-like core, depending on the opacity of the protoplanetary gas and on the growth rate of the core. We found that photo-evaporation produced negligible effects even at early ages (3–10 Myr). Transmission spectroscopy with JWST is very promising, and expected to provide constraints on atmospheric metallicity, abundance of H$_2$O, CO$_2$, CH$_4$ molecules, and to probe the presence of refractory elements. }  
   {TOI-837 offers valuable prospects for follow-up observations, which are needed for a thorough characterisation. JWST will help to better constraining the formation and evolution history of the system, and understand whether TOI-837\,$b$ is a Saturn-analogue. }


   \titlerunning{A characterisation study of the infant planetary system TOI-837}
    \authorrunning{Damasso et al.}

   \maketitle
%

\section{Introduction}
Although extremely important for understanding the formation and evolution of planets, the number of confirmed transiting exoplanets orbiting stars in stellar clusters and associations remains small so far. Only a few tens of planets have been found transiting stars younger than 100 Myr\footnote{We will refer to these as ``infant'' planets throughout the paper.} so far (e.g. \citealt{2016Natur.534..658D,David_2019ApJ...885L..12D,2020Natur.582..497P,2022A&A...667L..14Z}). Their detection is mainly hampered by the difficulties of observing transits in the light curves of stars located in dense stellar environments such as star clusters, although in recent years several techniques have been developed that minimise the problems due to contamination of nearby stars and the number of false positive detections (e.g. \citealt{2019MNRAS.490.3806N,2020MNRAS.495.4924N,2021MNRAS.505.3767N,2022MNRAS.511.4285B}). Moreover, the stellar activity signals prevailing in the photometric and spectroscopic time series of young stars make the detection and the characterisation of exoplanets difficult, although in the last few years giant strides were made in developing techniques to disentangle and filter out stellar activity signals from those due to planetary companions.

The advantage of investigating stellar clusters and associations is that, by using theoretical or empirical models, it is possible to derive precise stellar parameters such as radius, mass, effective temperature, chemical content, and age of the cluster members. Consequently, this in principle allows to measure precise age, mass and radius of the hosted exoplanets. At the same time, young stellar clusters offer the unique opportunity to investigate how the properties of the exoplanetary systems change with time right after their formation. Observing young systems at different ages is key to understand the timing of different physical processes occurring within the first $\sim$100 Myr after planet formation, which heavily influence the physical properties and architectures that we observe in mature planetary systems. These processes include planet-disc interactions (e.g. \citealt{2004ASPC..323..339L}), or planet-planet and planet-planetesimal interactions (e.g. \citealt{2008ApJ...686..580C,2011ApJ...742...72N}). Other less catastrophic processes affecting young short-period planets include thermal contraction and atmospheric mass-loss, driven either by core-powered mechanisms (e.g. \citealt{2018MNRAS.476..759G,2019MNRAS.487...24G}), or photo-evaporation (e.g. \citealt{2013ApJ...775..105O,2013ApJ...776....2L}). In particular, photo-evaporation is most active within the first $\sim$100 Myr, while core-powered mass-loss acts over longer timescales.

The detection and characterisation of infant exoplanets will help constraining the time scales of these mechanisms, including planet migration and tidal orbital circularisation, taking advantage of a detailed knowledge of the environment where the systems formed (e.g. stellar number density, radiation field). Obtaining accurate and precise radius and mass measurements, which in turn allow measuring planets’ bulk densities, by combining photometric and radial velocity (RV) time series is crucial to improve theoretical models of planetary formation and evolution at the earliest stages. In this regard, transiting young planetary systems, such as those detected by the NASA Transiting Exoplanet Survey Satellite (TESS; \citealt{ricker2015JATIS...1a4003R}), are particularly valuable, and they have been feeding most of the RV follow-up campaigns so far. However, young and, especially, infant stars show high levels of magnetic activity that cause quasi-periodic RV variations up to levels of hundreds of \ms, which can severely dwarf planetary signals (see e.g. \citealt{2020A&A...642A.133D,2022NatAs...6..232S} as extreme examples). High levels of magnetic activity also hamper the detection and precise modeling of small-depth photometric transits. Significant progress has been made for optimising the whole analysis framework: testing different RV extraction tools (e.g. methods based on the spectral cross-correlation function and template matching); using sophisticated methods to filter out stellar activity signals from RVs time series (e.g. techniques based on Gaussian processes regression), jointly modeled with time series of spectroscopic activity diagnostics and photometric light curves; improved methods for extracting and filtering light curves of young stars (e.g. \citealt{2023A&A...672A.144C}). Nonetheless, currently most young planets have mass upper limits, and only a limited sample have measured masses.

TOI-837 is an infant and very active solar-type star, bona-fide member of the young open cluster IC 2602, with an age of 35$^{+11}_{-5}$ Myr and metallicity $[Fe/H]$=-0.069$\pm$0.042 \citep{2020AJ....160..239B}.
A planet candidate TOI-837.01 was discovered by TESS. The candidate was also identified in the independent analysis by \citet{2020MNRAS.495.4924N} and labelled as PATHOS-30. The candidate, which has a period of $\sim$8.32 d and a Saturn-like size, was validated as a planet (then labelled as TOI-837\,$b$) by \cite{2020AJ....160..239B}, who determined a 3$\sigma$ mass upper limit of 1.2 $M_{\rm Jup}$.

To our knowledge, TOI-837\,$b$ is the youngest transiting validated planet known in an relatively young open cluster so far. It joined a small ensemble of transiting planets in young associations with well known age as AU Mic $b$ and $c$ in the $\beta$ Pic moving group (see e.g. \citealt{2020Natur.582..497P,Martioli2021,zicher22}); the multi-planet system V1298\,Tau in the Taurus-Ext association (see e.g. \citealt{David_2019ApJ...885L..12D,2022NatAs...6..232S,Feinstein_2022ApJ...925L...2F}); HIP\,67522\,$b$ \citep{Rizzuto_2020}, HD\,114082\,$b$ \citep{2022A&A...667L..14Z}, K2-33\,$b$ \citep{2016AJ....152...61M,2016Natur.534..658D}, and TOI-1227\,$b$ \citep{2022AJ....163..156M} in the Sco-Cen group; DS\,Tuc\,A\,$b$ in the Tuc-Hor association \citep{2019A&A...630A..81B,Newton_2019}.
The measurement of the mass through RV monitoring is of special interest, first for a final confirmation of the planetary nature of the stellar companion, but especially to infer its bulk density, internal and atmospheric structure, and study its formation and evolutionary history at early ages. The identification of the planet's Doppler signature in the RVs would also allow for the characterisation of the orbit of TOI-837\,$b$, providing insights on the formation and migration pathway at a such young age within a crowded environment which is potentially favourable to produce dynamical perturbations. 

In this paper, we present an updated analysis and characterisation of the system, based on photometric data that include recent TESS sectors, and archival HARPS high-resolution spectra (Sect. \ref{sec:data}). Fundamental stellar parameters are presented in Sect. \ref{sec:stellarparam}, while in Sect. \ref{sec:dataanalysis} we discuss the analysis of the photometric and spectroscopic data, including the modeling of the dominant signals due stellar magnetic activity. The implications of a significant detection of the RV Doppler signal, and consequently of the mass measurement of TOI-837\,$b$, for the formation and atmospheric evolution history of the system are discussed in Sect. \ref{sec:discussion}, where we also present interesting perspectives for a follow-up with JWST to characterise the atmospheric composition of the planet. 

We point out that, while our analysis was in progress, a parallel work about TOI-837 was submitted and posted online as a pre-print \citep{2024arXiv240413750B}. We emphasise that our results were not influenced by that of \cite{2024arXiv240413750B}. Our study should to be considered as an independent and alternative work, although it is based on the same HARPS and TESS raw data.

\section{Description of the datasets}
\label{sec:data}
\subsection{Photometry}

\begin{figure}
    \centering
    \includegraphics[width=0.5\textwidth]{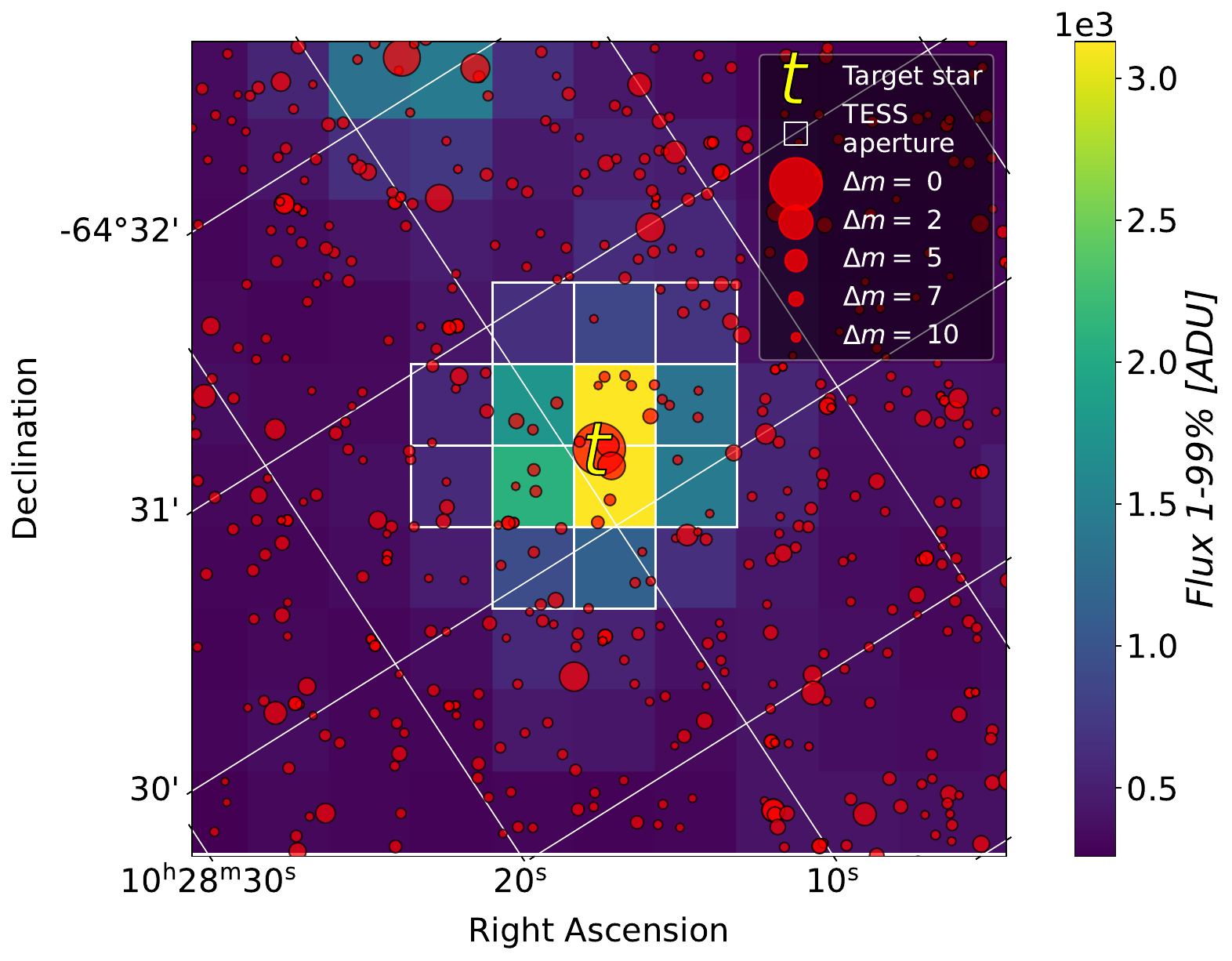}
    \caption{\textit{Gaia} DR3 stars identified in the TESS field of view of TOI-837, which is identified by the letter ‘t’. Several stellar contaminants fall within the TESS aperture. $\Delta m$ denotes the magnitude difference in the \textit{Gaia} G-band between each field star and TOI-837.}
    \label{fig:tessfov}
\end{figure}

TOI-837 has been observed by TESS in Sectors 10 and 11 (between 26 March and 21 May 2019), in Sectors 37 and 38 (between 2 April and 26 May 2021), and in Sectors 63 and 64 (between 10 March and 6 April 2023), spanning 749.3 days. We extracted the light curves from the short-cadence data by using the PATHOS pipeline described in \citet{2019MNRAS.490.3806N, 2020MNRAS.495.4924N, 2021MNRAS.505.3767N}. We corrected the light curves by adopting Cotrending Basis Vectors for short cadence data obtained and applied as described by \citet{2020MNRAS.498.5972N}. We did not make use of the PDCSAP (\citealt{2012PASP..124.1000S,2012PASP..124..985S,2014PASP..126..100S}) light curves because usually the light curves of these kind of highly variable stars are affected by systematic errors due to over-corrections in the official TESS pipeline (see \citealt{2022A&A...664A.163N} for details). Fig. \ref{fig:tessfov} shows the TESS field of view centred on TOI-837, revealing the presence of several photometric contaminants falling within the TESS aperture that contain the target, identified from \textit{Gaia} DR3 following the methodology outlined in \cite{2022MNRAS.516.4432M}. The stars nearby to TOI-837 and blended with it are accounted for to calculate the dilution factor, defined as the total flux from contaminant stars that fall into the photometric aperture divided by the flux of the target star. The dilution factor is used to correct the depth of the transit signal and, consequently, to provide a corrected measurement of the planet radius. For the case of TOI-837 we derived a significant dilution factor of 0.178$\pm$0.005.

\begin{figure*}
  \centering
  \includegraphics[width=0.9\textwidth,bb=15 200 600 712]{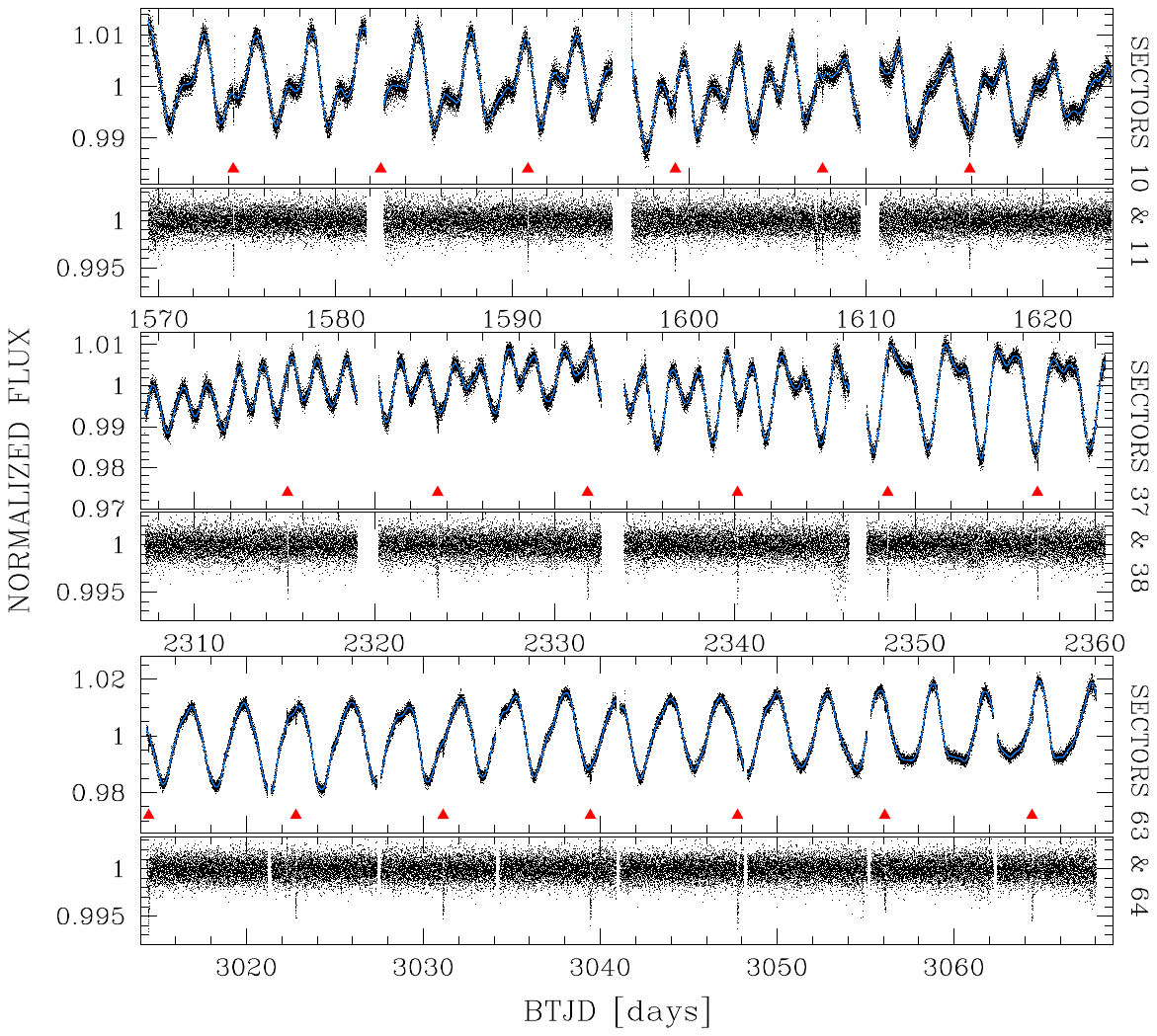}
  \caption{ The short cadence light curves of TOI-837 observed by TESS in Sectors 10 and 11 (top panel), 37 and 38 (middle panel), and 63 and 64 (bottom panel). For each pair of sectors, we report the short-cadence light curve and the detrending model (in azure), and the corresponding flattened light curve. The transits of TOI-837~$b$ are indicated with red triangles.  \label{fig:lc}}
\end{figure*}

We detrended the light curve by using the cosine estimator implemented in \texttt{w\=otan} package \citep{2019AJ....158..143H} with a window length of 0.25~day, and masking the transits by using the ephemeris reported by \citet{2020AJ....160..239B}. In Fig.~\ref{fig:lc} we show the light curve together with the model adopted for the flattening of the time series, and the flattened light curve. 

\subsection{Spectroscopy}
We downloaded 78 public HARPS spectra from the European Southern Observatory (ESO) archive, processed with the standard Data Reduction Software (DRS) v3.8 (template mask G2). The spectra were collected within program ID 110.241K.001/0110.C-4341(A) (Precise masses of very young transiting planets with HARPS; PI Yu). We discarded two spectra at epochs BJD 2459858.875731 and 2459880.857826 which, based on the information in the fits header, very likely correspond to a wrong source. For our analysis, we selected spectra with signal-to-noise ratio S/N$>$25, measured at the echelle order 50 ($\sim$5700\,\AA), resulting in 70 good epochs covering a time span of 101 days. The exposure times are 900 s and 1800 s. Data are provided in Table \ref{table:dataHS}.
The star is a fast rotator ($v\sin i_\star\sim$16 \kms), and the archival RVs calculated by default from the DRS cross-correlation function (CCF) are not trustworthy because of the narrow half-window used to calculate the CCF. Therefore, we do not analyse the DRS RVs in this study, instead we extracted the RVs using the \texttt{Python}-based code \texttt{SERVAL} (\texttt{spectrum radial velocity analyser}, version dated on 26 January 2022; \citealt{2018A&A...609A..12Z}), which adopts a procedure based on template-matching to derive relative RVs\footnote{The code is publicly available at \url{https://github.com/mzechmeister/serval}. We used the command line \textit{-safemode 2 -niter 4 -snmin 25 -ofac 0.30 -vrange 30} to process the HARPS spectra with SERVAL. We adopted a decreased oversampling factor for the spectra co-adding (i.e. ofac=0.3) as suggested by \cite{2018A&A...609A..12Z} to obtain a smoother template in case of noisy observations or fast rotators.}. The resulting RV time series is characterised by an RMS of 112 \ms and median internal error $\sigma_{\rm RV}$=10.2 \ms.    

\section{Stellar characterisation} 
\label{sec:stellarparam}

\subsection{Fundamental stellar parameters}
We derived the effective temperature ($T_{\rm eff}$), surface gravity ($\log g$), and iron abundance ([Fe/H]) using the spectral synthesis method to the co-added spectrum of the target. In particular, we used the spectral analysis tool \texttt{iSpec} (\citealt{blancocuaresma2014}, \citealt{blancocuaresma2019}) to measure these parameters. Specifically we considered the sixth version of the GES atomic line list (\citealt{Heiteretal2021}), both the MOOG (\citealt{Sneden1973}, version 2019) and SME (\citealt{ValentiPiskunov1996}, version 4.23) radiative transfer codes, and the MARCS (\citealt{Gustafssonetal2008}) and ATLAS9 (\citealt{castellikurucz2003}) grids of model atmospheres, obtaining consistent results. Regions encompassing the wing segments of the H$\alpha$, H$\beta$, and \ion{Mg}{i} triplet lines, together with \ion{Fe}{i} and \ion{Fe}{ii} lines in the 476-678\,nm spectral region were considered to constraints parameters. We then employed the non-linear least-squares Levenberg-Marquardt fitting algorithm (\citealt{Markwardt2009}) to iteratively minimise the $\chi^2$ value between the synthetic and observed spectra. Final mean values of the derived spectroscopic atmospheric parameters are listed in Table\,\ref{t:star_param}. As by-products, we also derived the projected rotational velocity ($v \sin i$), macroturbulence and microturbulence velocity ($V_{\rm macro}$, $V_{\rm micro}$). We also measured $T_{\rm eff}$ considering the line-depth ratio (LDR) method and appropriate LDR-$T_{\rm eff}$ calibrations developed at the same resolution as HARPS-N (see \citealt{Biazzoetal2011}), obtaining similar results within the uncertainties. 

\begin{figure}
    \centering
    \includegraphics[width=0.5\textwidth]{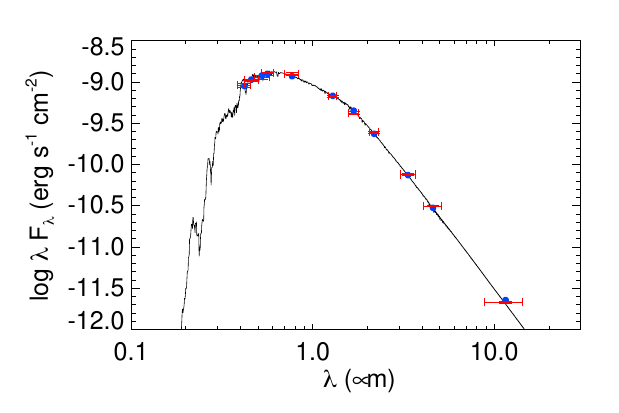}
    \caption{Spectral energy distribution of TOI-837 with the best-fit model over-plotted (solid line). Red and blue points correspond to the observed and predicted values, respectively.}
    \label{fig:sed}
\end{figure}

To determine the stellar physical parameters, namely mass, radius, and age, we simultaneously modelled the stellar Spectral Energy Distribution (SED; see Table~\ref{t:star_param} for the magnitudes used, and Fig.~\ref{fig:sed}) and the MIST stellar evolutionary tracks \citep{2015ApJS..220...15P} through a Bayesian differential evolution Markov chain Monte Carlo framework with the \texttt{EXOFASTv2} tool \citep{2017ascl.soft10003E, 2019arXiv190709480E}. We imposed Gaussian priors on the previously derived $T_{\rm eff}$ and [Fe/H], the \textit{Gaia} DR3 parallax, and the stellar age $\mathcal{N}(35,15)$~Myr. We also employed the PARSEC \citep{2012MNRAS.427..127B} stellar models instead of the MIST ones, and found practically equal stellar parameters and uncertainties. The derived stellar parameters are given in Table~\ref{t:star_param}.

Finally, we measured the equivalent width of the lithium line at $\sim$6707.8\,\AA, together with the Li abundance corrected for NLTE effects (\citealt{lindetal2009}). Their values, i.e. $EW_{\rm Li}=169\pm6$\,m\AA\,and $\log$A(Li)$^{\rm NLTE}$=$3.18\pm0.04$\,dex, are compatible with the membership of TOI-837 to the IC\,2602 young open cluster see, e.g., (\citealt{SestitoRandich2005, Jeffriesetal2023}).

All stellar parameters derived by our analysis are in agreement with those reported by \cite{2020AJ....160..239B} by less than 1$\sigma$. 

\begin{figure}
  \includegraphics[width=0.4\textwidth,bb=15 151 269 711]{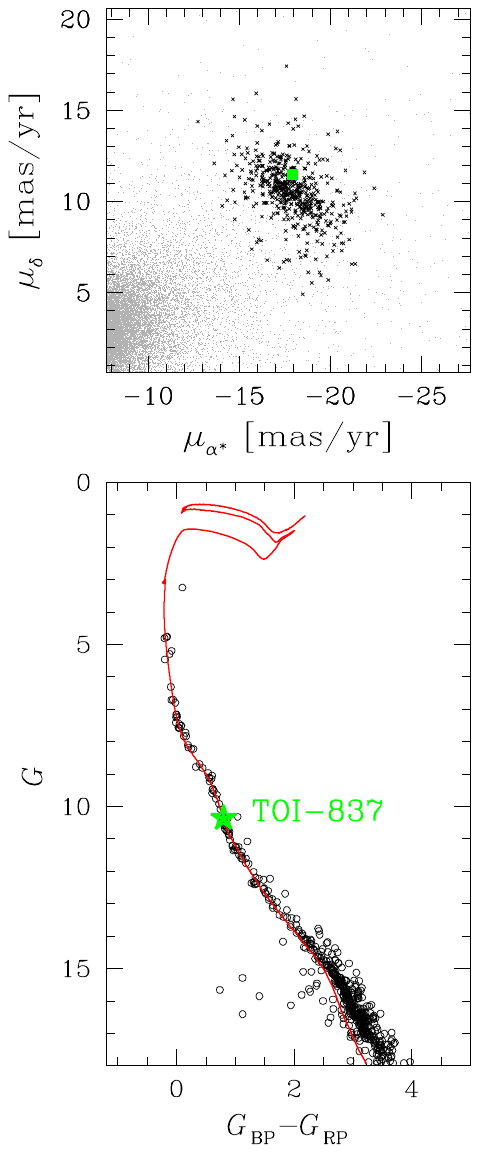}
  \caption{Proper motions and colour-magnitude diagram of the members of the open cluster IC~2602. Top panel shows the proper motions of the cluster members (in black) after a selection based on the parallax and the proper motions themselves. Bottom panel shows the \textit{Gaia} G-mag versus BP-RP colour-magnitude diagram of the same stars with superposed a BASTI-IAC isochrone of 35~Myr. In both panels, green symbols represent the target of this study. \label{fig:cmd}}
\end{figure}

Figure~\ref{fig:cmd} shows the \textit{Gaia} DR3 colour-magnitude diagram of IC~2602 members with over-imposed a 35~Myr BASTI-IAC isochrone\footnote{\url{http://basti-iac.oa-abruzzo.inaf.it/isocs.html}} \citep{2018ApJ...856..125H, 2021ApJ...908..102P}. The star is located on the main-sequence of the cluster and share the same proper motion and parallax. From the isochrone fitting, TOI-837 results to have a mass of about 1.12~M$_{\odot}$, a solar radius, and a $T_{\rm eff}\sim 6040$~K, in agreement with the spectroscopic stellar parameters adopted in this work.

\begin{table}[!htb]
   \caption[]{Stellar parameters of 
TOI-837 (CPD-63 1435)}
     \label{t:star_param}
     \tiny
     \centering
       \begin{tabular}{lcc}
         \hline
         \noalign{\smallskip}
         Parameter   &  Value & Ref  \\
         \noalign{\smallskip}
         \hline
         \noalign{\smallskip}
$\alpha$ (J2000)          &  10:28:08.99  &  \citep{gaia2016,gaia2021,gaia2023}  \\
$\delta$ (J2000)          &  -64:30:18.94   &  \citep{gaia2016,gaia2021,gaia2023}  \\
$\mu_{\alpha}$ (mas/yr)  &    -17.912$\pm$0.014  &  \citep{gaia2016,gaia2021,gaia2023}  \\
$\mu_{\delta}$ (mas/yr)  &     11.490$\pm$0.014  &  \citep{gaia2016,gaia2021,gaia2023} \\
$\pi$  (mas)             &    $7.0108\pm0.0124$ &  \citep{gaia2016,gaia2021,gaia2023}  \\
\noalign{\medskip}
B$_T$ (mag)                &    $11.12\pm0.06$  & \citep{2000}  \\
V$_T$ (mag)                  &    $10.64\pm0.05$     & \citep{2000}  \\
B$_{\rm Johnson}$ (mag)  & $11.107\pm0.036$ & \citep{2015AAS...22533616H} \\
V$_{\rm Johnson}$ (mag)  & $10.481\pm0.044$ & \citep{2015AAS...22533616H} \\
G (mag)                  &    $10.360\pm0.003$  &  \citep{gaia2016,gaia2021,gaia2023}  \\
i'$_{\rm Sloan}$ (mag)   & $10.138\pm0.042$   &  \citep{2015AAS...22533616H} \\
J$_{\rm 2MASS}$ (mag)    &   $9.392\pm0.030$  & \citep{cutri2003}   \\
H$_{\rm 2MASS}$ (mag)    &   $9.108\pm0.038$  & \citep{cutri2003}  \\
K$_{\rm 2MASS}$ (mag)    &   $8.933\pm0.026$  & \citep{cutri2003}  \\
WISE1 (mag)              &   $8.901\pm0.023$  & \citep{cutri2013} \\
WISE2 (mag)              &   $8.875\pm0.021$  & \citep{cutri2013} \\
WISE3 (mag)              &   $8.875\pm0.020$  & \citep{cutri2013} \\
$\rm A_{V}$ (mag)              &   $0.21\pm0.11$  & This paper (SED) \\
\noalign{\medskip}
T$_{\rm eff}$ (K)        &  6000$\pm$60        & This paper \\  
$\log g$                 &  4.60$\pm$0.20    & This paper  \\ 
& & (spectral analysis)\\
$\log g$                 &  4.453$^{+0.023}_{-0.029}$  & This paper  \\ 
& & (SED and MIST evolutionary tracks) \\ 
${\rm [Fe/H]}$ (dex)     &  -0.05$\pm$0.05      & This paper  \\ 
$S_{\rm MW}$             &   $0.414\pm0.014$    & This paper \\
$\log R^{'}_{\rm HK}$    &  -$4.311\pm0.018$ &  This paper \\  
$v\sin{i_{\star}}$ (km/s)      &  16.5$\pm$1.3  & This paper  \\  
${\rm P_{\rm rot,\,\star}}$ (d)  &   2.973$\pm$0.006  &   This paper  \\
& & (TESS photometry) \\
$EW_{\rm Li}$ (m\AA)     &     $169\pm6$ &  This paper   \\
A(Li)                    &     3.18$\pm$0.04  &  This paper   \\
Mass (M$_{\odot}$)       &   $1.109^{+0.038}_{-0.048 }$    & This paper  \\
Radius (R$_{\odot}$)     &    $1.036\pm0.015$    & This paper  \\
Density ($\rho_{\odot}$)    &    $1.08^{+0.04}_{-0.04}$ & This paper \\
Luminosity (L$_{\odot}$) &    1.27$^{+0.10}_{-0.09}$    & This paper  \\
Age (Myr)               &    $35_{-5}^{+11}$ & \cite{2020AJ....160..239B}  \\
Age (Myr)               &    $39_{-12}^{+14}$ & This paper \\
& & (SED and MIST evolutionary tracks)\tablefootmark{a} \\
         \noalign{\smallskip}
         \hline
      \end{tabular}
      \tablefoot{\tiny
				\tablefoottext{a}{Posterior derived after using the age interval estimated by \cite{2020AJ....160..239B} as a prior. }
			}
\end{table}

\subsection{A stellar companion at wide separation}
\label{sec:bin}

A star at about 2.3" from TOI-837 (namely TIC 847769574) was previously mentioned by \citet{2020AJ....160..239B}. 
They reported a similar parallax and proper motion from \textit{Gaia} DR2, but slightly discrepant values of the distance, to infer that this object is a member of the IC 2602 open cluster but not physically bound to TOI-837.
We reconsider the physical association of the two stars exploiting \textit{Gaia} DR3 astrometry. Their parallax are now compatible at 1.6 $\sigma$ level. Therefore, the argument of different distances no longer applies. Proper motion differ by a significant amount in declination (2.0 mas/yr) while it agrees within error bars in right ascension.
This difference corresponds to a velocity difference of about 1.33 \kms at the distance of TOI-837. This is fully compatible with orbital motion in a bound pair with separation of $\sim$ 330 au.
From the density of targets observed around TOI-837 with astrometric properties compatible with cluster membership in \textit{Gaia} DR3, a low probability of the order of 0.1$\%$ is derived for catching another cluster object at a projected separation as small as 2.3". We conclude that the characteristics of the object found at close separation from TOI-837 are fully compatible with those of a gravitationally bound object, and that the presence of an unbound cluster member projected at such close separation is very unlikely.
We estimate M=0.33 \msun for the mass of the stellar companion using the stellar models by \cite{baraffe2015}.
With a magnitude difference of 4.91 in \textit{Gaia} G-band, larger in V-band, the impact of the bound companion on the time variations of the RVs data of TOI-837 discussed below is expected to be negligible.


\section{Data analysis} \label{sec:dataanalysis}

\subsection{Photometric stellar rotation period} \label{sec:rotper}
We performed a Gaussian process (GP) regression analysis to model each pair of TESS sectors individually, with the goal of measuring the stellar rotation period $P_{\rm rot,\star}$, and characterising the evolutionary time-scales of the periodic variations as seen in the light curve. We adopted the GP quasi-periodic (QP) kernel, whose covariance matrix is defined as (e.g. \citealt{haywood2014}) 
		\begin{gather} 
			\label{eq:eqgpqpkernel}
			k_{QP}(t, t^{\prime}) = h^2\cdot\exp\Bigg[-\frac{(t-t^{\prime})^2}{2\lambda^2} - \frac{\sin^{2}\Bigg(\pi(t-t^{\prime})/\theta\Bigg)}{2w^2}\Bigg] + \nonumber \\
			+\, (\sigma^{2}(t)+\sigma^{2}_{\rm jit})\cdot\delta_{t, t^{\prime}}
		\end{gather}
where: $t$ and $t^{\prime}$ are two different epochs of observations; $\sigma(t)$ is the uncertainty of a data point at epoch $t$; $\delta_{t, t^{\prime}}$ is the Kronecker delta; $\sigma_{\rm jit}$ is a constant jitter term added in quadrature to the formal uncertainties $\sigma(t)$ to account for other sources of uncorrelated noise. The GP hyper-parameters are $h$, which denotes the scale amplitude of the correlated signal (uniform prior: $\mathcal{U}$(0,5)); $\theta$, which corresponds to the stellar rotation period $P_{\rm rot,\star}$ (uniform prior: $\mathcal{U}$(0,1) days); $w$, which describes the harmonic complexity of the rotation period within a complete stellar rotation cycle (uniform prior: $\mathcal{U}$(0,1)); $\lambda$, that represents the decay timescale of the correlations, and is related to the temporal evolution of the magnetically active regions (uniform prior: $\mathcal{U}$(0,1000) days). We included a linear trend into the model for sectors 63 and 64. To perform the GP regression we used the publicly available \textsc{python} module \textsc{george} v0.2.1 \citep{2015ITPAM..38..252A}. We explored the full parameter space using the Monte Carlo (MC) nested sampler \texttt{MultiNest v3.10} (e.g. \citealt{Feroz2019}), through the \texttt{pyMultiNest} wrapper \citep{Buchner2014}. 
The weighted mean for $P_{\rm rot,\star}$ from the three measurements is 2.973$\pm$0.006 d. For the $\lambda$ hyper-parameter the weighted mean is 4.83$\pm$0.05 d, implying the presence of active regions that evolve quickly on a time scale shorter than two rotation cycles.   

\subsection{Frequency content of radial velocities and spectroscopic activity diagnostics} \label{sec:chromindex}

\begin{figure}[h!]
    \centering
    \includegraphics[width=0.45\textwidth]{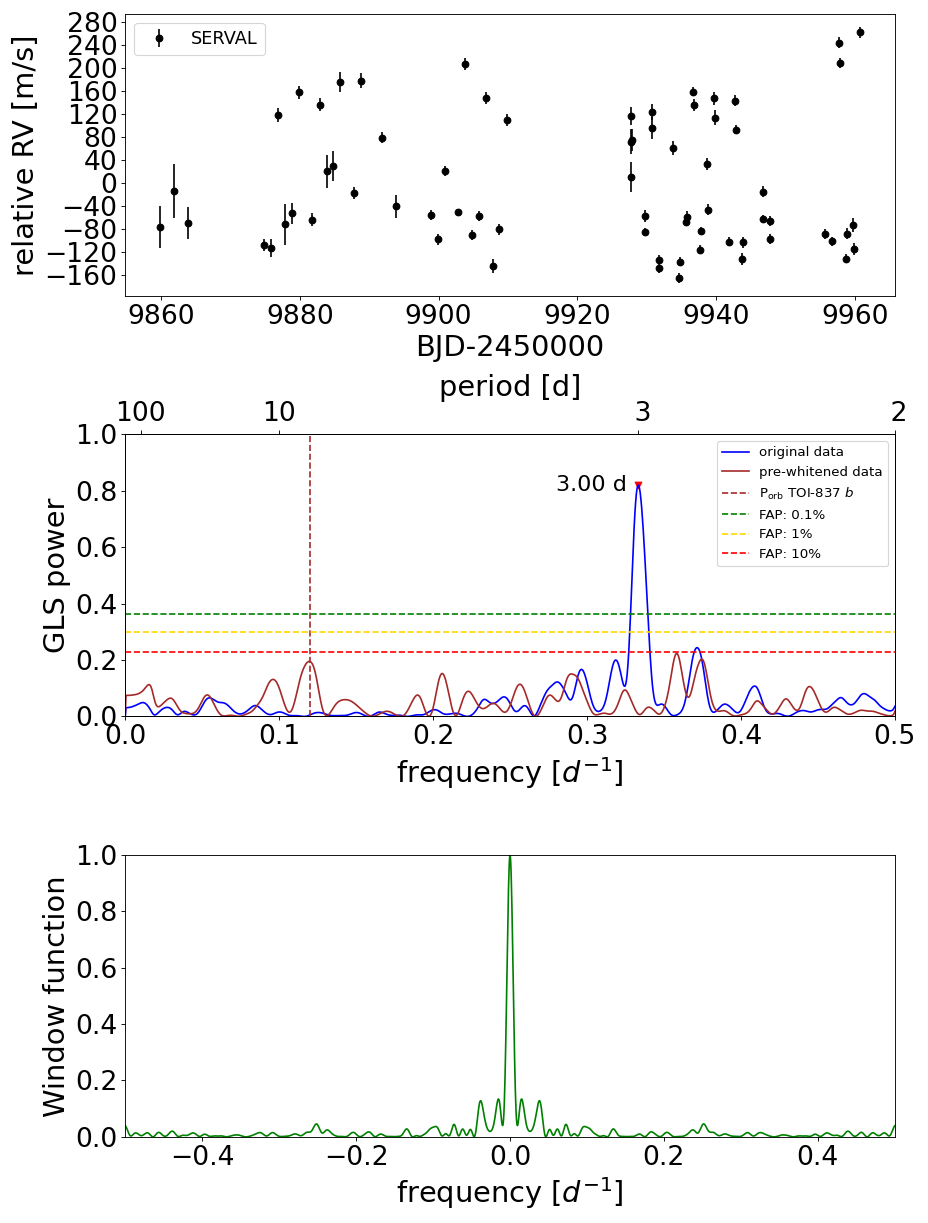}
    \caption{\textit{Upper panel.} RV time series of TOI-837 (SERVAL extraction). \textit{Middle panel.} GLS periodogram of the orginal and pre-whithened data. False alarm probability (FAP) levels for the periodogram of the original RV dataset are indicated by horizontal dashed lines. \textit{Lower panel.} Window function of the data.}
    \label{fig:rvgls}
\end{figure}

Fig. \ref{fig:rvgls} shows the RV time series and the corresponding Generalised Lomb-Scargle (GLS; \citealt{2009A&A...496..577Z}) periodogram. A very significant peak is found at the stellar rotation period $P_{\rm rot,\, \star}$, showing that the observed RV scatter is mainly due to activity. Interestingly, the periodogram of the pre-whitened data (i.e. after removing the best-fit sinusoid calculated by GLS) shows a peak very close to the orbital period of planet $b$.

\begin{figure*}[h!]
    \centering
    \includegraphics[width=0.49\textwidth,trim={0cm 0 0.cm 0cm}]{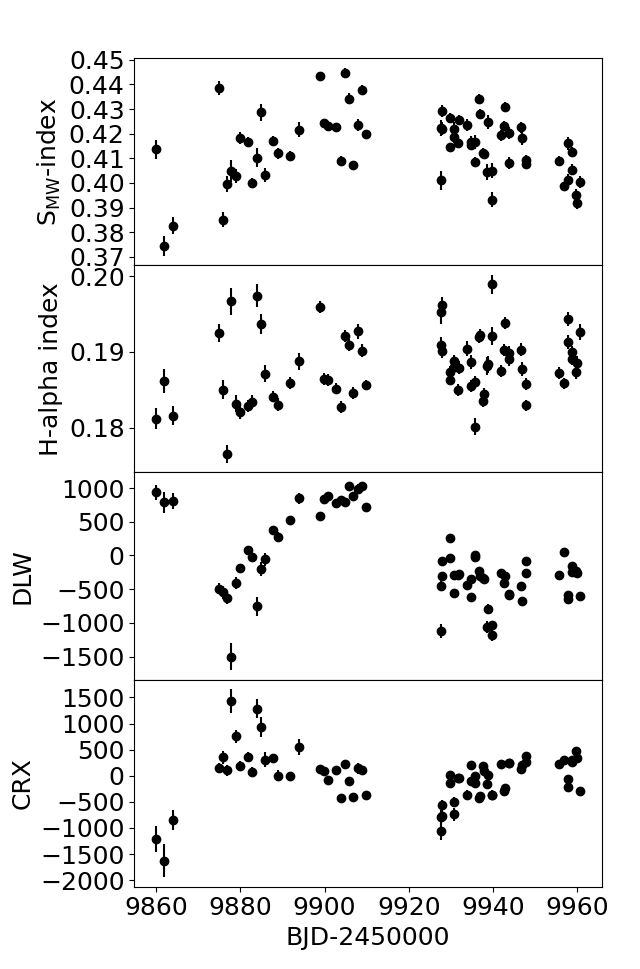}
    \includegraphics[width=0.49\textwidth,trim={0.cm 0.cm 0cm 0cm}]{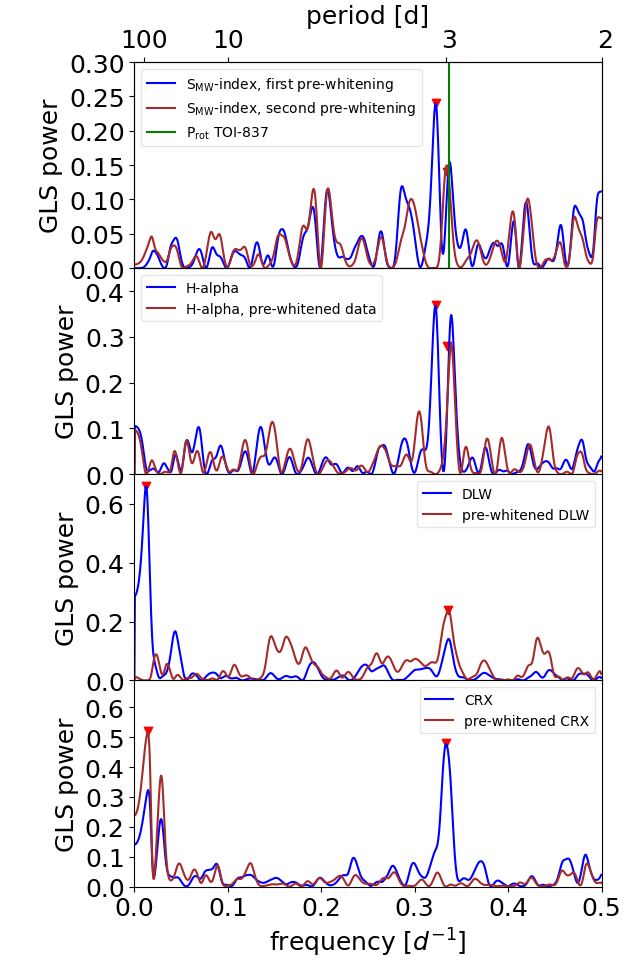}
    \caption{\textit{Left panel.} Time series (black dots) of activity diagnostics extracted from HARPS spectra. \textit{Right panel.} GLS periodograms of the time series, and pre-whitened time series, shown in the left panel. The vertical green line identifies the stellar rotation period measured from the TESS light curve.}
    \label{fig:actindper}
\end{figure*}

We analysed the time series of spectroscopic activity diagnostics that can be used to disentagle planetary signals identified in the RVs from those due to stellar activity (Table \ref{table:actdiagnostics}). Using the code \texttt{ACTIN2} \citep{2018JOSS....3..667G,2021A&A...646A..77G}, we calculated the chromospheric activity index S$_{\rm MW}$, derived from the CaII H\&K line doublet and calibrated to the Mount Wilson scale following \cite{2021A&A...646A..77G}, and the activity diagnostic derived from the H-alpha line. Additionally, we calculated the activity diagnostics differential line width (DLW) and the chromatic index (CRX) using \texttt{SERVAL} \citep{2018A&A...609A..12Z}. DLW is equivalent to the full-width half maximum of the spectral cross-correlation function, and the CRX, defined as the slope of the slope in a plot RVs vs. logarithm of the wavelengths of the individual echelle orders, allows to investigate how much the RVs are ``chromatic'' due to effects of activity. Time series and periodograms are shown in Fig. \ref{fig:actindper}. The S-index shows a long-term modulation, for which it is not possible determine a periodicity because of the limited time span of the data. After correcting the data fitting a quadratic trend, the GLS periodogram of the residual time series shows the main peak at 3.1 days. After pre-whitening the residuals (i.e. removing a sinusoid with period of 3.1 days), the periodogram of the new residual time series has the main peak at 2.99 days. The time series of the H-alpha index does not show a long-term trend, but the periodogram has similarities with that of the pre-whitened S-index. It shows two peaks close to the photometric $P_{\rm rot,\star}$, with the one with higher power at 3.1 days. The periodogram of the pre-whitened data has the main peak at 2.95 days. The conclusion from the analysis of the S$_{\rm MW}$ and H-alpha indexes is that there are two distinct sinusoidal-like signals with periodicities very close to the photometric $P_{\rm rot,\star}$, possibly indicative of differential rotation.  
The DLW time series shows a clear long-term modulation with a significant GLS periodicity of $\sim81$ days. The periodogram of the pre-whitened time series shows a single peak related to $P_{\rm rot,\star}$. The periodogram of the CRX index is dominated by a single peak related to $P_{\rm rot,\star}$, while that of the pre-whitened CRX data shows a peak at $\sim$70 days. Thus, the frequency analysis of both the DLW and CRX indexes reveals the presence of a modulation over a time scale much longer than the rotation period.  

We employed the \texttt{Python} code \texttt{pyrhk}\footnote{\url{https://github.com/gomesdasilva/pyrhk/blob/master/pyrhk.py}} to convert the values of the S$_{\rm MW}$ index to the standard $\log R'_{\rm HK}$ chromospheric emission ratio using the standard formula proposed by \citet{Noyes+1984}, and bolometric correction taken from \cite{1982A&A...113....1M}. We obtained $\log R^{\prime}_{\rm HK} = -4.311\pm 0.018$. This measurement was employed to estimated also the coronal X-ray luminosity of TOI-837, that we need to study the photo-evaporation history of the planet (Sect.\ \ref{sec:photoevap}). 

\subsection{RV and photometry joint analysis} \label{sec:rvphotoanalysis}

We performed a RV and photometry joint analysis to derive orbital and physical parameters of TOI-837 $b$, and search for undetected companions. From the detrended light curve we removed data points corresponding to flares which are clearly visible in Fig. \ref{fig:lc}.

The MC \texttt{MultiNest} set-up for the joint RV and photometric analysis included 500 live points, a sampling efficiency of 0.3, and a Bayesian evidence tolerance of 0.5. We used the code \texttt{batman} \citep{Kreidberg2015} for modeling the photometric transits. Concerning the light curve modeling, we used the stellar density $\rho_{*}$ as a free parameter from which we derived the $a_{\rm b}/R_{\star}$ ratios at each step of the MC sampling (e.g. \citealt{2007ApJ...664.1190S}), using a Gaussian prior based on the mass and radius derived in Sect. \ref{sec:stellarparam}. We adopted a quadratic law for the limb darkening, and fitted the coefficients $u_{\rm 1}$ and $u_{\rm 2}$ using the parametrization and the uniform priors for the coefficients q$_1$ and q$_2$ given by \cite{kipping2010} (see Eq. 15 and 16 therein). We also introduced a constant jitter $\sigma_{\rm jit}$ added in quadrature to the nominal photometric uncertainties.

\begin{figure*}[h!]
    \centering
    \includegraphics[width=0.9\linewidth]{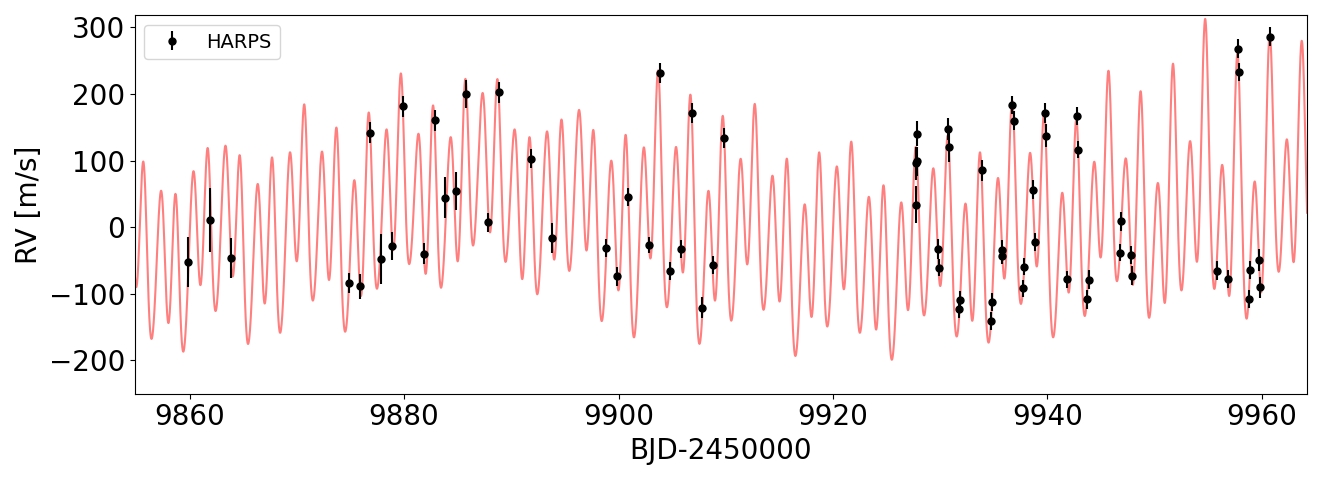}\\
    \includegraphics[width=0.45\linewidth]{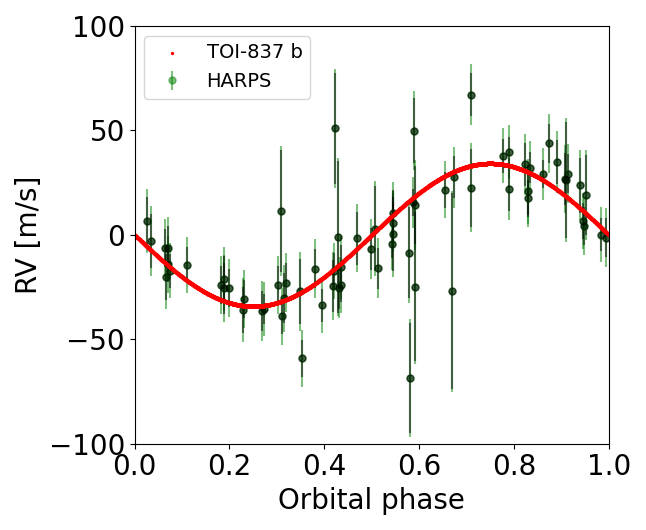}
    \includegraphics[width=0.45\linewidth]{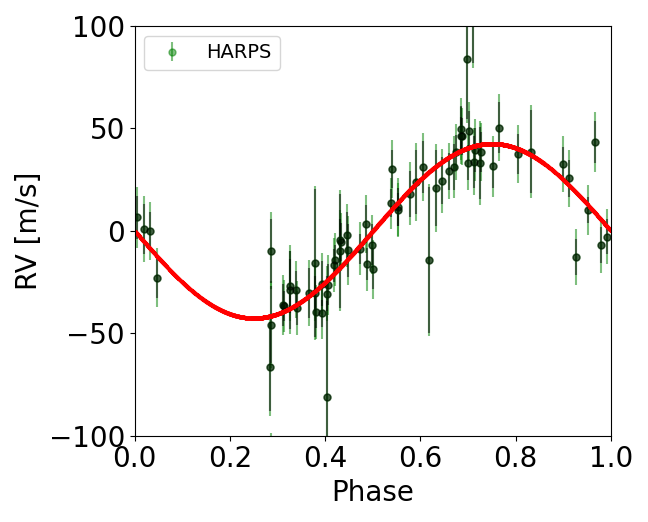}\\
    \includegraphics[width=0.45\linewidth]{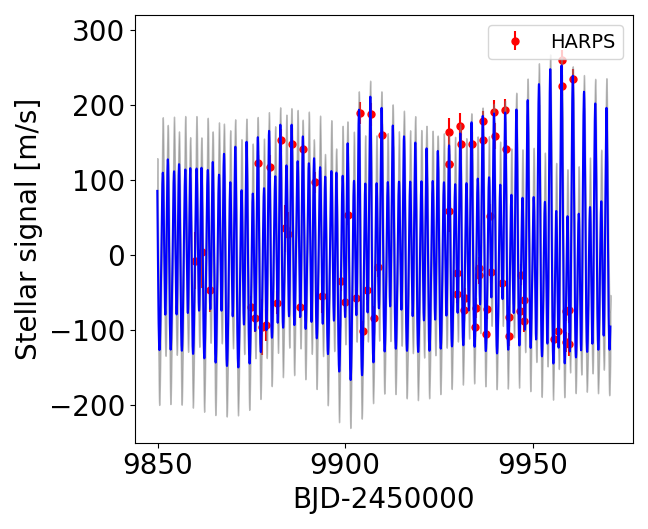}
    \includegraphics[width=0.45\linewidth]{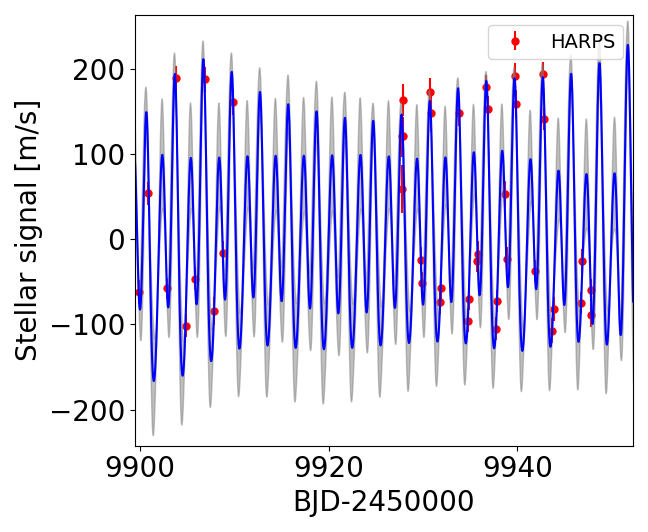}\\
    \caption{Best-fit models related to the RV dataset obtained from the joint RV+light curve analysis described in Sect. \ref{sec:rvphotoanalysis}).\textit{ First row.} RV time series with over-plotted the adopted best-fit model (red curve; activity plus Doppler signal due to TOI-837\,$b$).    
    \textit{Second row.} Phase-folded Doppler RV signal of TOI-837 due to planet $b$ (left panel), and the signal with period $\sim$ 74 days that we interpret as due to stellar activity (right panel). Nominal error bars are indicated in black, while the error bars with an uncorrelated jitter term added in quadrature are indicated in green.
    \textit{Third row.} Stellar activity correlated signal in the RV time series (red dots) as fitted by a GP regression including the sum of two quasi-periodic kernels. The blue line represents the best-fit model, and the 1$\sigma$ confidence interval is shown as a shaded gray area. The panel on the right is a zoomed-in view of the left panel, for better readability.
    To calculate all the theoretical curves shown here, we adopted the median values of the model parameters in Table \ref{tab:resultjointfit}.
    }
    \label{fig:rvfitplot}
\end{figure*}

\begin{figure}[h!]
    \centering
    \includegraphics[width=0.5\textwidth]{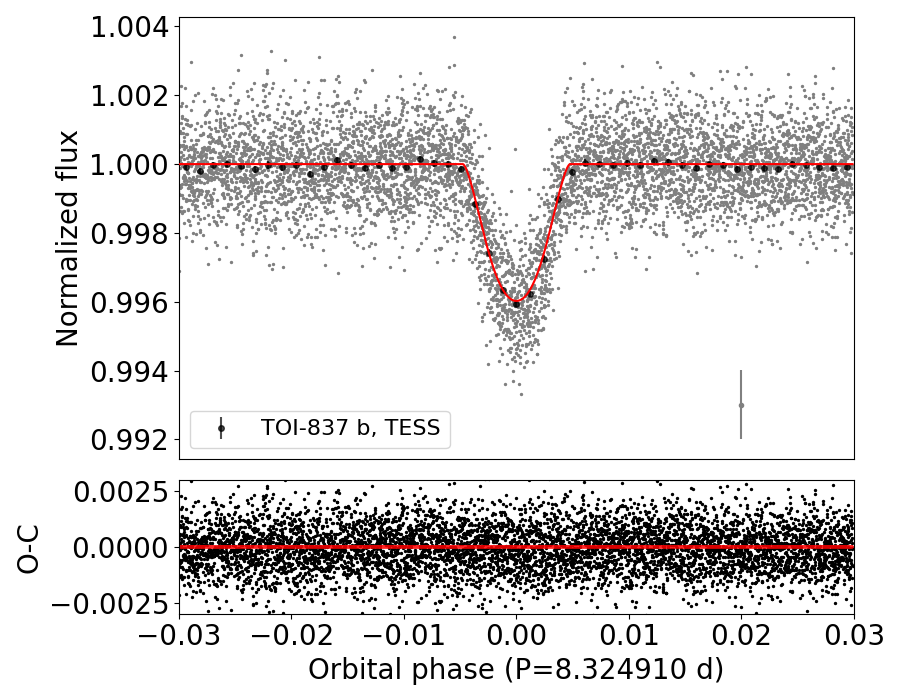}
    \caption{Phase-folded transit light curve of TOI-837 $b$, and O-C residuals. The data point on the bottom right of the first panel shows the median error bar of the complete photometric time series. }
    \label{fig:transit}
\end{figure}

Concerning the fit of the RVs, we considered a base model that includes a Keplerian to fit the Doppler signal of the transiting planet --either keeping the eccentricity $e_b$ as a free parameter, or fixing $e_b=0$-- an offset $\gamma_{\rm RV}$, and a secular acceleration $\dot{\gamma}$. The dominant signal due to the stellar activity, which is modulated over $P_{\rm rot,\star}$, was modelled using a GP regression, adopting the quasi-periodic kernel of Eq. \ref{eq:eqgpqpkernel}. For the base model, we found that assuming a circular orbit for TOI-837 $b$ is statistically favoured over the eccentric case, and we adopt $e_b=0$ for the rest of our analysis. More complex RV models were tested to search for additional significant signals present in the dataset, and possibly ascribable to planetary companions. Periods of the additional signals were sampled uniformly either in the range 10 to 100 days or below 7 days. We found that the best-fit model for our RV data includes a sinusoid to fit a signal with period longer than the orbital period $P_b$ of TOI-837 $b$, and a second GP quasi-periodic kernel, which is added to the first kernel, to fit a third signal with a period shorter than $P_b$. This model is statistically strongly favoured over the base model, with $\Delta \ln \mathcal{Z}_{\rm 3\,signals-1\,planet}=+6.0$, where $\mathcal{Z}$ denotes the Bayesian evidence calculated by \texttt{MultiNest}. We adopt this as our reference model. Priors and results for both the base and adopted model are summarised in Table \ref{tab:resultjointfit}, where the parameters of the long-period sinusoid are labelled with the subscript ``long-period'', while the hyper-parameters of the second GP quasi-periodic signal are labelled with the subscript ``2''. The posteriors are shown in the corner plot of Fig. \ref{fig:cornerplot}. We provide a description and interpretation of the results in the following. The most relevant plots are shown in Fig. \ref{fig:rvfitplot} and \ref{fig:transit}, for the RVs and photometry respectively. 

Analysing the results for the RVs, the Doppler signal due to TOI-837\,$b$ is robustly detected ($K_b$=34.2$^{+4.9}_{-5.3}$ \ms, $\sim$6.5$\sigma$ significance level), and translates into a true mass $m_b$=116$^{+17}_{-18}$ \mearth taking into account the measured orbital inclination angle $i_b$=86.96$\pm$0.05 degrees from the transit photometry. The sum of two GP quasi-periodic kernels is able to effectively filter-out a correlated signal linked to $P_{\rm rot,\star}$. This activity-related signal is composed of a dominant component ($h_1$=181$^{+68}_{-58}$ \ms) with a well-constrained period $\theta_1$=3.000$^{+0.003}_{-0.004}$ d, and a secondary component ($h_2$=36$^{+10}_{-7}$ \ms) with period $\theta_2$=3.06$^{+0.25}_{-0.28}$ d and a shorter evolutionary time scale ($\lambda_2$=6$^{+16}_{-3}$ d), which is similar to the values measured for the $\lambda$ hyper-parameter from the TESS light curve (Sect. \ref{sec:rotper}). We found evidence in the periodogram of the $S_{\rm MW}$ and H-alpha activity indicators for a similar bi-modal periodicity related to $P_{\rm rot,\star}$ (Sect. \ref{sec:chromindex}), which supports the use of a sum of two quasi-periodic kernels.  
The second long-period sinusoid included into our best-fit model is significant (semi-amplitude $K_{\rm long\,period}$=42.6$^{+11.2}_{-12.2}$ \ms), with periodicity $P_{\rm long\,period}$=74.8$^{+13.4}_{-9.1}$ d. Very likely, this signal is not due to a second companion in the system, and it appears related to stellar activity, although the interpretation of the physical mechanism responsible for that is not straightforward. A high positive correlation exists between this long-period RV signal and the pre-whitened CRX index (i.e. after removing the signal due to stellar rotation; Fig. \ref{fig:rvcrxcorr}), with a correlation coefficient $\rho_{Pearson}$=+0.71. An RV Doppler shift induced by a planet is achromatic, i.e. its properties are independent from the wavelength of the stellar spectral lines, while an anti-correlation with the CRX is observed for RV signals that are related to spot-dominated stellar activity \citep{2018A&A...609A..12Z}. The observed positive correlation for TOI-837 does not have an immediate astrophysical explanation, but it prevents to interpret the RV signal as due to an additional companion in the system. 
We note that the values of the parameters of planet TOI-837\,$b$ are in agreement within the error bars for the two models shown in Table \ref{tab:resultjointfit}. This means that the characterisation of planet $b$ is not affected by a more complex modeling of the activity-related signals in the RV time series, although this complexity is required by the data.  

The best-fit model for the light curve (Fig. \ref{fig:transit}) corresponds to a grazing transit (impact parameter $b$=0.91$\pm$0.02) and to a planetary radius $r_b$=9.71$^{+0.93}_{-0.60}$ \rearth. 
Our measured mass and radius imply a most-likely Saturn-like bulk density for TOI-837\,$b$, $\rho_b$=0.68$^{+0.20}_{-0.18}$ \gcm ($\rho_{\rm Sat}$=0.687 \gcm).

As a final note, we found that running \texttt{TLS} on the TESS light curve after masking the transits of planet TOI-837 $b$ does not result in a significant detection of additional signals.

	\begin{table*}[h!]
		\caption{Priors and best-fit values of the free and derived parameters related to the joint RV+light curve modeling.}
		\label{tab:resultjointfit}
		\small
		\begin{center}
			\begin{tabular}{llcc}
				\hline
				\textbf{Parameter}   & \textbf{Priors} & \multicolumn{2}{c}{\textbf{Best-fit value}\tablefootmark{a}}\\
                    & & GP + 1 sinusoid  & Two GPs + 2 sinusoids  \\
                    & & (base model) & (adopted model) \\ 
				\hline
				\noalign{\smallskip} 
				\textit{Free parameters:} & \\
				\noalign{\smallskip}
                    $h_1$ [m/s] & $\mathcal{U}$(0,300) &  $138^{+56}_{-32}$ & $181^{+68}_{-58}$ \\
                    \noalign{\smallskip}
				$\lambda_1$ [d] & $\mathcal{U}$(0,1000) &  $32^{+12}_{-10}$  & $400^{\rm +334}_{\rm -207}$ \\
				\noalign{\smallskip}
				$w_1$ & $\mathcal{U}$(0,1) &  $0.33^{+0.17}_{-0.11}$  & $0.46^{+0.16}_{-0.14}$ \\ 
				\noalign{\smallskip}
				$\theta_1$ [d] & $\mathcal{U}$(0,5) & $3.01\pm0.02$ & $3.000^{+0.003}_{-0.004}$ \\      
				\noalign{\smallskip}
                    $h_2$ [m/s] & $\mathcal{U}$(0,300) & - & $36^{+10}_{-7}$ \\
                    \noalign{\smallskip}
				$\lambda_2$ [d] & $\mathcal{U}$(0,1000) &  -  & $6^{\rm +16}_{\rm -3}$ \\
				\noalign{\smallskip}
				$w_2$ & $\mathcal{U}$(0,1) &  -  & $0.39^{+0.18}_{-0.13}$ \\ 
				\noalign{\smallskip}
				$\theta_2$ [d] & $\mathcal{U}$(0,5) & - & $3.06^{+0.25}_{-0.28}$ \\      
				\noalign{\smallskip}
				$K_b$ [\ms] & $\mathcal{U}$(0,100) & $32.9^{+5.8}_{-5.9}$ & $34.2^{+4.9}_{-5.3}$\\
				\noalign{\smallskip} 
				orbital period, $P_b$ [d] & $\mathcal{U}$(8.31,8.32) & $8.3249104\pm0.0000040$ & $8.3249102\pm0.0000037$ \\
				\noalign{\smallskip}
				time of inferior conjunction, T$_{conj,\,b}$ [BJD-2450000] & $\mathcal{U}$(9290.2,9290.3) & $9290.2151\pm0.00030$ & $9290.2151\pm0.00027$ \\
				\noalign{\smallskip}
                $K_{\rm long\,period}$ [\ms] & $\mathcal{U}$(0,100) & - & $42.6^{+11.2}_{-12.2}$\\
				\noalign{\smallskip} 
				period, $P_{\rm long\,period}$ [d] & $\mathcal{U}$(10,100) & - & 74.8$_{-9.1}^{+13.4}$ \\
				\noalign{\smallskip}
			     T$_{\rm RV=0,\,long\,period}$ [BJD-2450000] & $\mathcal{U}$(9860,9970) & - & $9906.4^{+3.4}_{-3.0}$ \\
				\noalign{\smallskip}
    			acceleration, $\dot{\gamma}$ [\ms d$^{-1}$] & $\mathcal{U}$(-1,1) & $-0.22^{+0.68}_{-0.53}$ & $-0.11\pm0.47$ \\
       			\noalign{\smallskip}
                    $R_{\rm b}$/$R_{\rm \star}$  & $\mathcal{U}(0.0, 0.3)$ & $0.086^{+0.008}_{-0.006}$ & $0.079^{+0.008}_{-0.005}$ \\ 
                    \noalign{\smallskip}
                    $i_{\rm b}$ [degrees] & $\mathcal{U}(80, 90)$ & $86.96\pm0.06$ & $86.96\pm0.05$  \\
                   \noalign{\smallskip}
                    limb dark. parametrization, $q_{\rm 1,\,TESS}$ & $\mathcal{U}(0, 1)$ &  $0.57^{+0.25}_{-0.21}$ &  $0.55^{+0.23}_{-0.20}$ \\
                    \noalign{\smallskip}
                    limb dark. parametrization, $q_{\rm 2,\,TESS}$ & $\mathcal{U}(0, 1)$ &  $0.45^{+0.33}_{-0.28}$ &  $0.46^{+0.32}_{-0.29}$ \\
                    \noalign{\smallskip}
                    $\rho_{\rm \star}$ $[\rm \rho_{\odot}]$ & $\mathcal{G}(1.08, 0.04)$ & $1.08\pm0.04$ & $1.08\pm0.03$ \\
                    \noalign{\smallskip}
				$\sigma_{\rm jit\,RV,\: HARPS}$ [\ms] & $\mathcal{U}(0,50)$ & $25\pm4$ & $10.5^{+10.4}_{-6.7}$ \\ 
				\noalign{\smallskip}
				$\gamma_{\rm RV,\, HARPS}$ [\ms] & $\mathcal{U}(-200,200)$ & $53^{+51}_{-58}$ & $44^{+83}_{-94}$ \\
				\noalign{\smallskip}
                    $\sigma_{\rm jit\,TESS}$ [ppm] & $\mathcal{U}(0,0.1)$ & $0.00030\pm0.00003$ & $0.00030\pm0.00003$ \\ 
				\noalign{\smallskip}
				$\gamma_{\rm TESS}$ [ppm] & $\mathcal{U}(-0.01,0.01)$ & $-0.00005\pm0.00001$ & $-0.00005\pm0.00001$\\
				\noalign{\smallskip}
                    \textit{Derived parameters}\\
                    \noalign{\smallskip}
                    eccentricity $e_{\rm b}$ & - & 0 (fixed) & 0 (fixed) \\
                    \noalign{\smallskip}
                     periastron argument, $\omega_{\rm \star,\, b}$ ~[rad] & - & $\pi$/2 (fixed) & $\pi$/2 (fixed) \\
                    \noalign{\smallskip}
                    limb dark. coeff., $u_{\rm 1,\,TESS}$ & - &  $0.66^{+0.42}_{-0.41}$ & $0.65\pm0.40$ \\
                    \noalign{\smallskip}
                    limb dark. coeff., $u_{\rm 2,\,TESS}$ & - & $0.07^{+0.44}_{-0.45}$ & $0.06^{+0.45}_{-0.42}$ \\
                    \noalign{\smallskip}
                    semi-major axis, $a_b$ [au] & - & $0.083\pm0.001$ & $0.083\pm0.001$ \\
                    \noalign{\smallskip}
                    impact parameter, $b$ & - & $0.91\pm0.03$ & $0.91\pm0.02$\\
                    \noalign{\smallskip}
                    transit duration [days] & - & $0.0904^{+0.0072}_{-0.0076}$& $0.0905^{+0.0075}_{-0.0068}$ \\
				\noalign{\smallskip}
                    radius, $r_b$ [\rearth] & - &  $9.74^{+0.96}_{-0.67}$ &  $9.71^{+0.93}_{-0.60}$ \\
                    \noalign{\smallskip}
                    mass, $m_b$ [\mearth] & - & $112\pm20$ & $116^{+17}_{-18}$ \\
                    \noalign{\smallskip}    
                    bulk density, $\rho_b$ [\gcm] & - & $0.64^{+0.22}_{-0.19}$  & $0.68^{+0.20}_{-0.18}$ \\
                    \noalign{\smallskip}
                    equilibrium temperature\tablefootmark{b}, $T_{\rm eq.,\,b}$ [K] & - & \multicolumn{2}{c}{1022$\pm$13} \\
                    \noalign{\smallskip}
				\hline
                    \noalign{\smallskip} 
                    Bayesian evidence, $\ln \mathcal{Z}$ & & 34450.3 & 34456.3 \\
				\noalign{\smallskip} 
                    \hline
			\end{tabular}
			\tablefoot{\tiny
				\tablefoottext{a}{The best-fit values, the lower and upper uncertainty are given as the $50^{\rm th}$, $16^{\rm th}$ and $84^{\rm th}$ percentiles of the posterior distributions, respectively.}
                \tablefoottext{b}{Assuming zero albedo.}
			}
		\end{center}
	\end{table*}

\section{Discussion} \label{sec:discussion}

\begin{figure}[h!]
    \centering
    \includegraphics[width=0.5\textwidth]{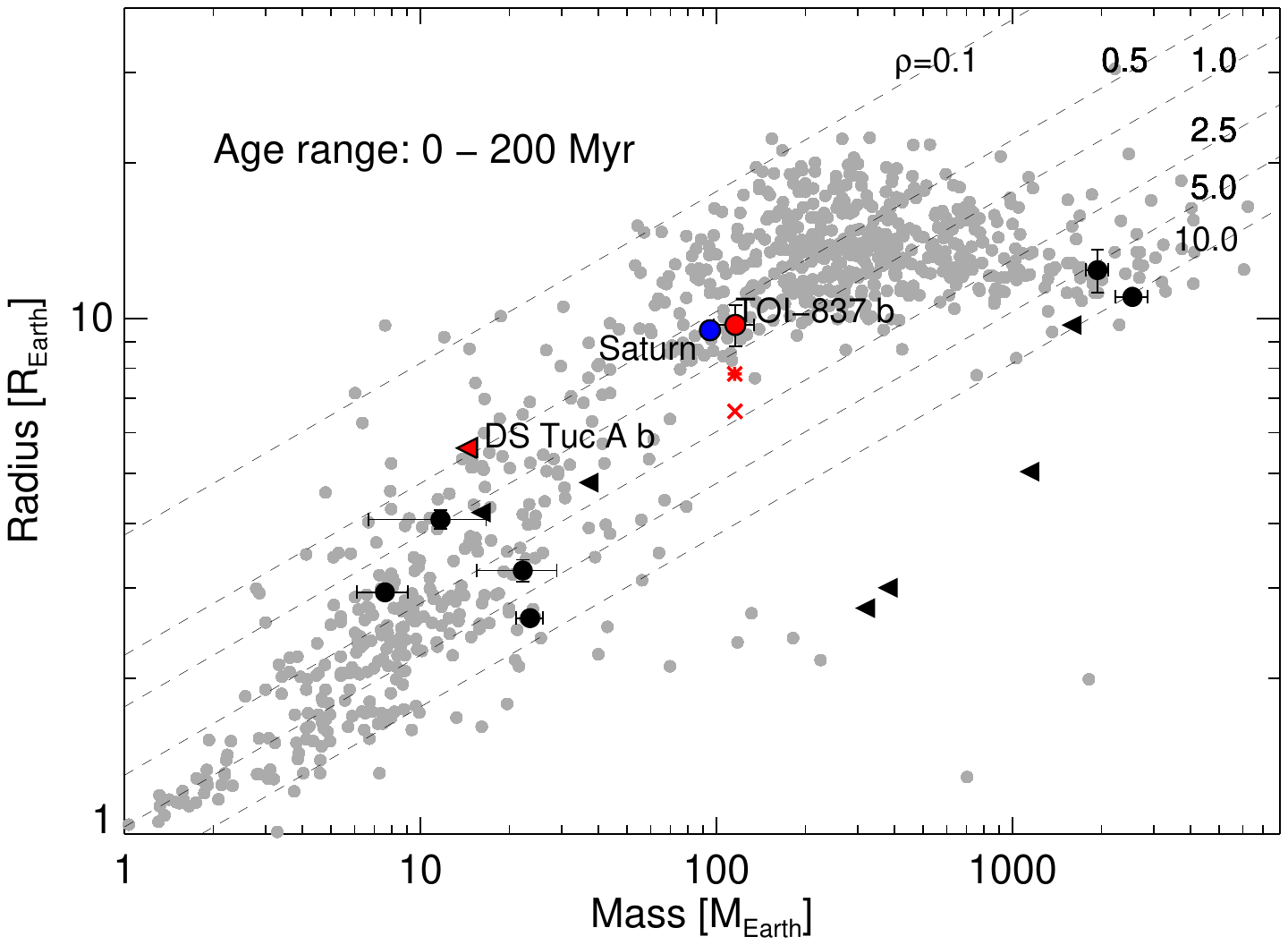}\\
    \includegraphics[width=0.5\textwidth]{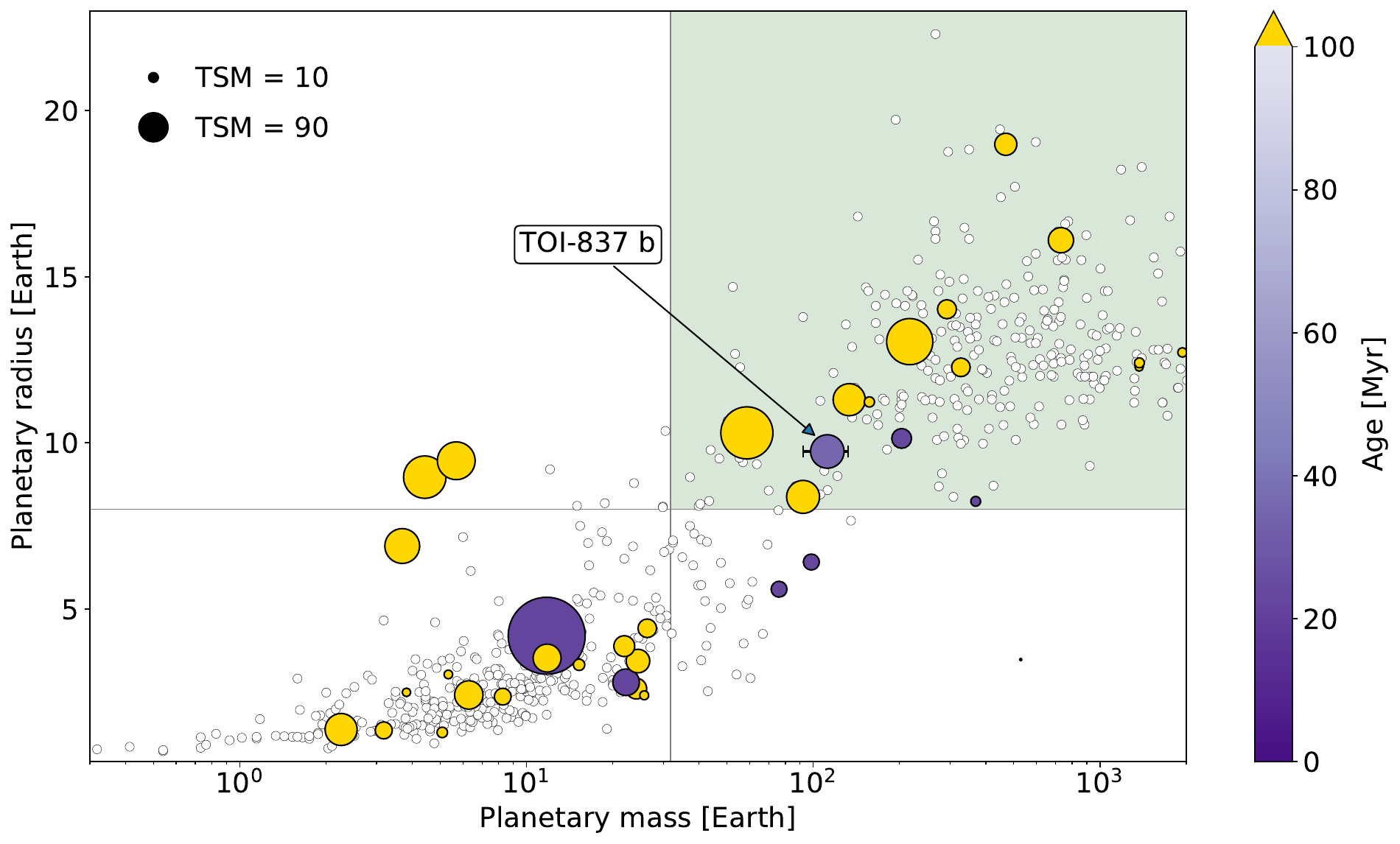}
    \caption{\textit{Upper panel.} Mass-radius diagrams showing transiting planets with age less than 200 Myr (black symbols), taking the age upper uncertainties into account for the definition of the age bin. Black dots indicate planets with a measured mass, while black triangles represent planets for which only a mass upper limit is available. Grey dots represent a sample of older planets with masses and radii known at least at 30 and 10\%, respectively. The current location of the planet TOI-837\,$b$ is indicated with a red dot, while the red asterisk and cross symbols denote the expected locations at an age of 1 Gyr for the cases \textit{rock-ice core/low atmospheric opacity} and \textit{rock-iron core/high atmospheric opacity}, respectively (see Table \ref{table:par} and Sect. \ref{sec:photoevap} for details). A red triangle indicates the location of the planet DS\,Tuc\,A\,$b$ (1$\sigma$ mass upper limit; \citealt{2021A&A...650A..66B}). Saturn is identified by a blue dot. Diagonal dashed lines indicate locations of planets with an equal density.
    \textit{Lower panel.} Mass-radius diagram of planets younger than 200 Myr, with the dot size representing the TSM metrics for prospects of spectroscopic follow-up with JWST. The mass of the planets is shown on a logarithmic scale. The giant planets ($R_{\rm p} > 8~R_{\oplus}$ and $M_{\rm p} > 0.1~M_{\rm J}$) are represented within the light green area.}
    \label{fig:mrdiagram}
\end{figure}

We show in Fig. \ref{fig:mrdiagram} the mass-radius diagram for transiting planets younger than 200 Myr discovered so far, which have measured masses (black circles), or just mass upper limits (black triangles)\footnote{The four confirmed planets of the 20 Myr old system V1298 Tau are not shown in the mass-radius diagram. The high complexity encountered so far in modeling and characterising this system resulted in quite different published values for the planets' masses, with no consensus reached yet. }. Their number is still too scarce to allow for demographics of infant/adolescent planets compared to mature planets (which are identified with grey dots in the plot), therefore it is important to increase the number of detection in the future, and provide robust constraints on the planetary masses, trying to overcome the limits imposed by stellar activity. That is essential to confirm whether highly irradiated, very young exoplanets show larger radii with respect to the mature counterpart with a similar mass, as both theory (e.g. \citealt{linder_2019,2020MNRAS.498.5030O}) and a few observational evidences currently suggest (e.g. \citealt{2020AJ....160..108B,2021A&A...650A..66B,2022AJ....163..156M}). If so, we should expect that, in general, infant/adolescent planets will follow evolutionary tracks on the mass-radius diagram as they age. We investigate the evolution of TOI-837\,$b$ on the mass-radius diagram in Sect. \ref{sec:photoevap}.

In the context of the known exoplanet population younger than 200 Myr, TOI-837\,$b$ joins a very small number of warm/hot giant-sized planets with a measured mass, namely HD\,114082\,$b$ \citep{2022A&A...667L..14Z}, Qatar-4\,$b$ \citep{Alsubai_2017}, with the exception of the 17 Myr old HIP\,67522\,$b$, for which a mass upper limit is available so far \citep{Rizzuto_2020}. The detection of short-period transiting Saturn/Jupiter-sized planets around very young stars should not be greatly hampered by stellar activity (even though the accuracy and precision of the radius measurements can be affected), therefore the impact of observational biases on the observed scarcity should be mild. If the investigation of a larger sample of young stars will confirm that the low occurrence rate is real, this will greatly help constraining the timescales of disk migration mechanisms for massive planets.

In Fig. \ref{fig:mrdiagram} we highlighted the planet DS\,Tuc\,A\,$b$ \citep{2019A&A...630A..81B,2021A&A...650A..66B} with a red triangle, because this system shares interesting similarities with TOI-837, which possibly justify a dedicated comparative study in the future. DS\,Tuc\,A is coeval to, but with a later spectral type than TOI-837, being a 40$\pm$5 Myr old G6V main component of a binary system. It has a rotation period and activity levels detected in the RVs which are very close to that of TOI-837, and as TOI-837 it hosts a planet with a similar semi-major axis ($\sim$0.08 au). The radius of DS\,Tuc\,A\,$b$ is about 0.5 $R_{\rm Jup}$, but so far only a conservative mass upper limit of 14.4 \mearth has been derived from a RV time series measured from HARPS spectra \citep{2019A&A...630A..81B,2021A&A...650A..66B}. DS\,Tuc\,A\,$b$ is then very likely an inflated super-Earth, at least one order of magnitude less massive than TOI-837\,$b$. Possibly, future RV follow-ups could pin down the mass, and allow for a characterisation of the internal structure and composition of DS\,Tuc\,A\,$b$. If the two planets shared a similar formation history within protoplanetary disks with similar properties, the characterisation of DS\,Tuc\,A\,$b$ might help to effectively reconstruct the formation history of TOI-837\,$b$. In the following, we present a preliminary investigation of the formation and evolutionary histories of TOI-837\,$b$, based on our measured planet's parameters, population synthesis simulations and models of mass-loss through photo-evaporation.

\subsection{Formation history}
We first estimated the timescale of tidal circularisation of the TOI-837\,$b$ orbit for a hypothetical eccentricity $e_{\rm b}=0.15$ from Eq.~1 in \citet{2008ApJ...686L..29M}, by assuming for both the host star and the planet a modified tidal quality factor of $Q'_{\rm p}=Q'_{\rm \star}\sim10^{6}$ \citep{2017A&A...602A.107B} as well as planet spin-orbit synchronisation. We found an orbit circularisation timescale close to the age of the universe. Therefore, the circular orbit observed for TOI-837\,$b$ is very likely primordial, and indicates that the planet has undergone a smooth disk migration, with no excitation of its eccentricity.

Building on this indication, we explored the possible formation tracks of the planet in the framework of the pebble accretion scenario using our Monte Carlo implementation of the Planet Growth and Migration (\texttt{GroMiT}) code\footnote{\url{https://doi.org/10.5281/zenodo.10593198}}. \texttt{GroMiT} uses the planetary core growth and migration treatments from \citet{johansen2019}, the pebble isolation mass scaling law from \citet{bitsch2018}, the gas accretion, gap-opening and gap-driven migration treatments from \citet{tanaka2020}, and incorporates the condensation sequence treatment in protoplanetary disks from \citet{turrini2023} to track the formation history of a planet from its original seed and formation region to its final mass and orbit, based on the characteristics of its native circumstellar disk. 

\begin{figure*}[h!]
\includegraphics[width=\textwidth]{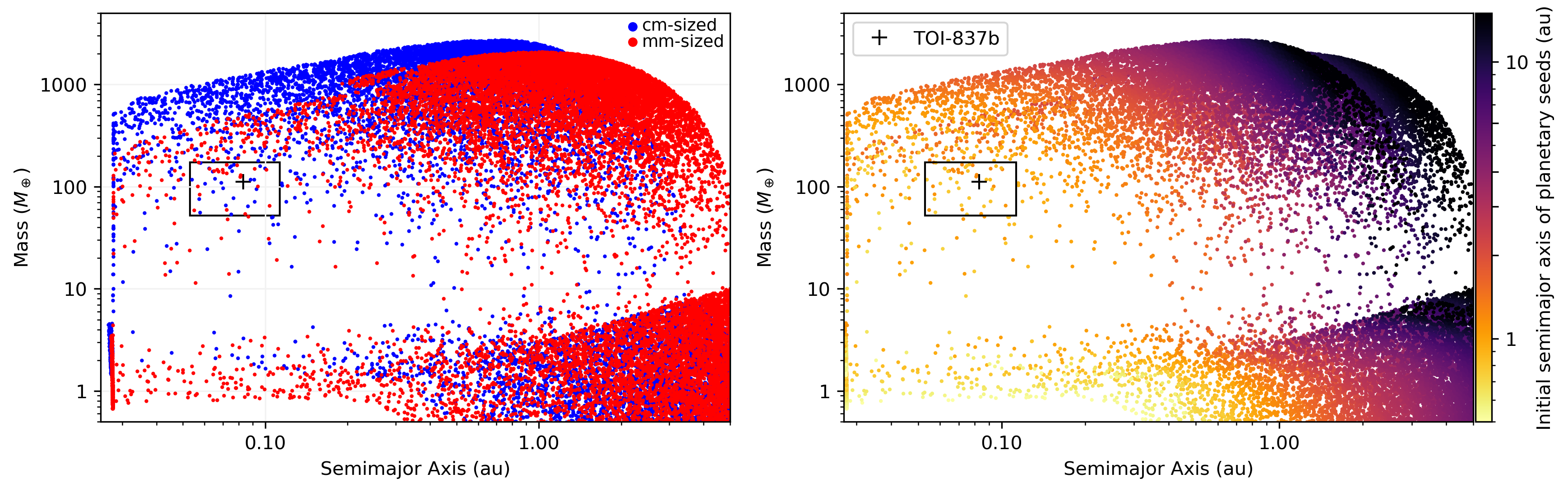}
\caption{Synthetic planetary populations resulting from the Monte Carlo runs of the \textit{GroMiT} code, plotted in the final mass--final semimajor axis space. The cross indicates the median mass and semimajor axis of TOI-837\,$b$ while the box shows the associated $3\sigma$ range of the planetary mass and semimajor axis. \textit{Left:} We plot here the results for both mm-sized (red) and cm-sized (blue) pebbles. 
\textit{Right:} The same as in the left panel, but including a colour scale to indicate the initial semimajor axes of the planetary seeds. }\label{fig:population-synthesis}
\end{figure*}

We consider a circumstellar disk with total mass of 0.06 M$_\odot$, characteristic radius of 50\,au, gas surface density at the characteristic radius of 4.2\,g/cm$^2$, temperature profile defined as $T=T_{0} \cdot r^{-0.5}$ and  temperature $T_{0}$ at 1\,au as 200\,K (see Table \ref{tab:simparam} and \citealt{Mantovan24b} for more details). In the population synthesis simulations with \textit{GroMiT} we considered both mm-sized and cm-sized pebbles, performing 10$^{5}$ Monte Carlo extractions for each pebble size. As shown in the left-hand panel of Fig. \ref{fig:population-synthesis}, the characteristics of the planet are reproduced with both selected pebble sizes. The major difference between the two populations is that in the case of mm-sized pebbles, the planetary seed must form within the first 2 Myr of the disk lifetime to successfully grow to TOI-837\,$b$'s mass. In the case of the cm-sized pebbles populated disk, on the other hand, the planetary seeds need to appear after 2.5 Myr in the disk lifetime to produce equivalent planets, otherwise the higher results in the growth to larger masses. As discussed below, in our template disk the planets that form in the innermost few au from the star are characterised by small pebble isolation masses. This translates into long envelope contraction times that delay the onset of the runaway gas accretion phase, limiting their final masses to a few earth masses. Planets that fulfil the conditions for entering the runaway gas growth phase can easily become more massive than TOI-837\,$b$ or migrate to the inner edge of the disk. As a consequence, the mass range between 3 and 100\,M$_\oplus$ is sparsely populated near the orbital location of TOI-837\,$b$.

\begin{figure}[h!]
\includegraphics[width=\columnwidth]{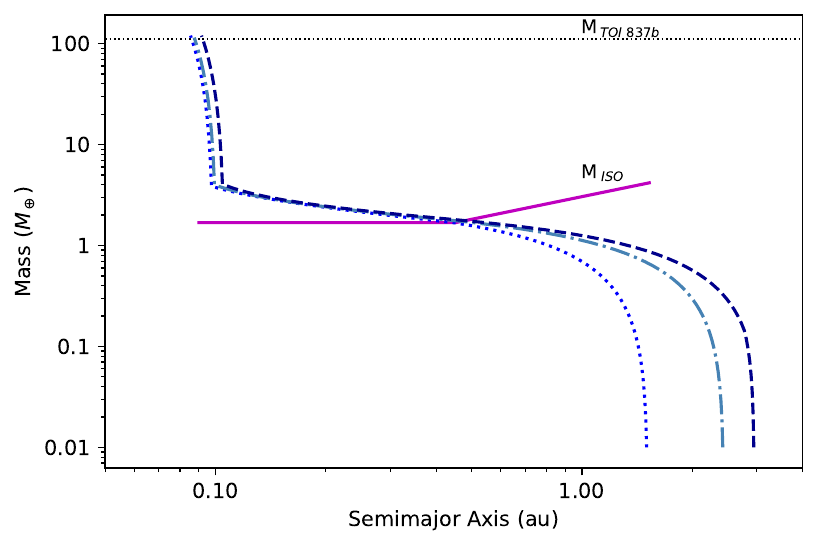}
\caption{Illustrative examples of planetary growth tracks falling within the 3$\sigma$ box for planet $b$. The growth tracks are projected in the semimajor axis--planetary mass space. All successful tracks are associated with formation regions comprised between 1 and 4\,a and have planetary cores of about 2\,M$_\oplus$. The core mass is identified by the mass value where the growth tracks intercept the pebble isolation mass curve (purple line).}\label{fig:growth-tracks}
\end{figure}

In the right-hand panel of Fig. \ref{fig:population-synthesis} we show the resulting population of synthetic planets plotted in the semimajor orbit versus planetary mass parameter space, zooming on the planets with final semimajor axis inwards of 4\,au. The colour scale indicates the initial semimajor axis of the planetary seeds, the black cross marks the observed mass of TOI-837\,$b$, while the black box represents the 3$\sigma$ uncertainty region associated with the measured mass and orbit of the planet. The solutions we find within this uncertainty region all point to the original seed having originated between 2 and 4\,au. As illustrated in Fig. \ref{fig:growth-tracks}, the core mass values of the compatible solutions are set by the pebble isolation mass to about 2\,M$_{\oplus}$ with the onset of the runaway gas accretion occurring at a fraction of au. Due to the range of possible formation distances of the planetary seed emerging from the population synthesis, both rock/metal cores and ice/rock mixtures are possible for the planetary core, although in the second case ice likely provides a limited fraction of the core mass. 

\begin{table}
    \caption{The input parameters used to run the MC modified \textit{GroMiT} code to produce simulated planets and investigate the formation history of TOI-837\,$b$.}
    \label{tab:popsythesis}
    \centering
    \begin{tabular}{l c c}
        \hline \hline
    \multicolumn{2}{c}{Simulation Parameters} \rule{0pt}{2.3ex} \rule[-1ex]{0pt}{0pt}\\
    \hline
      \multicolumn{1}{l}{N$^\circ$ of Monte Carlo runs} & \multicolumn{1}{c}{2$\times$10$^5$} \rule{0pt}{2.3ex} \rule[-1ex]{0pt}{0pt}\\
      \multicolumn{1}{l}{Seed formation time} & \multicolumn{1}{c}{0.1--4.0$\, \times \, 10^6$\,yr} \\
      \multicolumn{1}{l}{Disk lifetime} & \multicolumn{1}{c}{$5.0 \times 10^6$\,yr} \\
    \hline
    \multicolumn{2}{c}{Star, Planet \& Disk properties} \rule{0pt}{2.3ex} \rule[-1ex]{0pt}{0pt}\\
    \hline
      \multicolumn{1}{l}{Stellar Mass} & \multicolumn{1}{c}{1.109$\,$M${_\odot}$} \rule{0pt}{2.3ex} \rule[-1ex]{0pt}{0pt}\\
      \multicolumn{1}{l}{Disk Mass} & \multicolumn{1}{c}{0.06$\,$M${_\odot}$} \\
      \multicolumn{1}{l}{Disk characteristic radius R$_c$} & \multicolumn{1}{c}{50.0$\,$au} \\
      \multicolumn{1}{l}{Surface density @ R$_c$} & \multicolumn{1}{c}{4.2\,g cm$^{-2}$} \\
      \multicolumn{1}{l}{Temperature T$_0$ @ 1$\,$au} & \multicolumn{1}{c}{200$\,$K} \\
      \multicolumn{1}{l}{Disk accretion coefficient, $\alpha$} & \multicolumn{1}{c}{0.01}\\
      \multicolumn{1}{l}{Turbulent viscosity, $\alpha$$_\nu$}& \multicolumn{1}{c}{0.0001}\\
      \multicolumn{1}{l}{Pebble size} & \multicolumn{1}{c}{1$\,$mm--1$\,$cm}\\
      \multicolumn{1}{l}{Seed Mass} & \multicolumn{1}{c}{0.01$\,$M${_\oplus}$} \\
      \multicolumn{1}{l}{Initial envelope mass} & \multicolumn{1}{c}{0.0$\,$M${_\oplus}$} \\
      \multicolumn{1}{l}{Initial semimajor axis} & \multicolumn{1}{c}{0.5--20.0$\,$au} \\
    \hline
  \end{tabular}
  \label{tab:simparam}
\end{table}

To gain further insight on the formation history of TOI-837\,$b$, we coupled the population synthesis study with an investigation of the possible interior structures compatible with the planetary radius and mass emerging from the retrievals. To this end, we performed a Monte Carlo study based on the equations for the core and envelope radii from \cite{LopFor14}. We explored both solutions for both the solar opacity and enhanced opacity cases from \cite{LopFor14}, which are associated to faster and slower contraction timescales respectively. We performed $10^6$ MC runs where we extracted the stellar age and the planetary mass from the posterior distributions emerging from the retrievals using normal distributions truncated to zero, and the core mass fraction from a uniform distribution between 0 and 1. We used the resulting values to compute the planetary radius using Eq. 2 and 4 from \cite{LopFor14}, selecting only those combinations of values producing planetary radii, masses and densities compatible within $3\sigma$ with the modal values from the retrievals.

\begin{figure*}[h!]
\centering
\includegraphics[width=\textwidth]{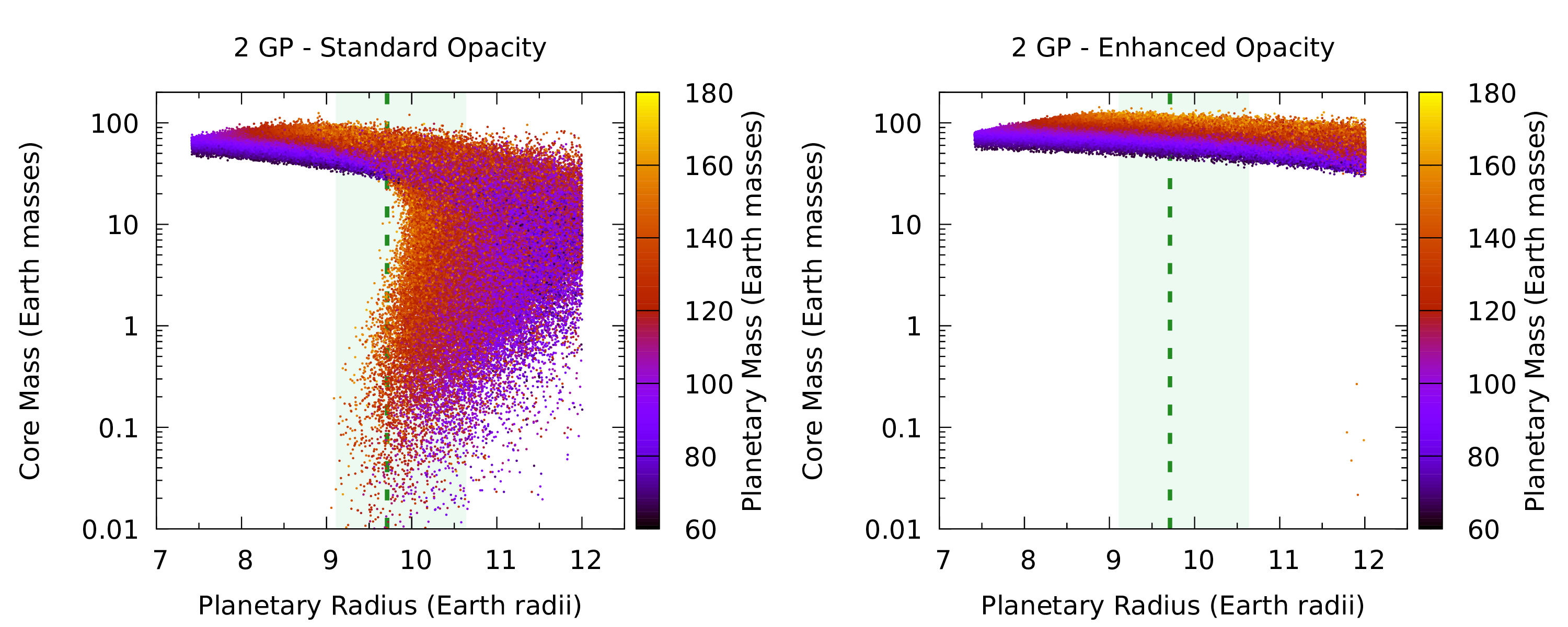}
\caption{Monte Carlo simulations of the interior structures that are compatible within $3\sigma$ with the observed planetary radius (dashed green vertical line; the shaded green area include radii within 1$\sigma$ of the best-fit value) and planetary mass in the solution with 2 GP quasi-periodic kernels, when their uncertainties and the uncertainty on the stellar age are taken into account. Fitting the planetary radius in the case of enhanced opacity of the envelope requires large cores in excess of 30 M$_\oplus$, while the faster contraction of envelopes with solar opacity allows for smaller cores compatible with the mass range estimated for Saturn.}\label{fig:interior-structure}
\end{figure*}

The resulting distributions of compatible interior structures are shown in Fig. \ref{fig:interior-structure}. As can be immediately seen, the slower contraction associated with envelopes with enhanced opacity strongly favours large cores with masses systematically greater than 30 M$_\oplus$ and generally well in excess of 50-60 M$_\oplus$ (see also Fig. \ref{fig:mrcore}). Solutions with smaller cores are sparse and are associated with planetary radii more than 2$\sigma$ larger than the modal value (see Fig. \ref{fig:interior-structure}). The faster contraction associated with envelopes of standard opacity allows instead for a second family of solutions with smaller core masses (Fig. \ref{fig:interior-structure}). This second family of solution spans cores of the order of that estimated for Saturn by gravity and ring seismology investigations (about 20 M$_\oplus$, \citealt{mankovich2021}) to those with very small cores (sub-Earth mass). Overall, we found that about 31-32\% of the solutions shown in Fig, \ref{fig:interior-structure} for the cases of standard opacity are synthetic planets with radius, mass and density within 1$\sigma$ of the retrieved best-fit values for TOI-837\,$b$, and about 1 out of 6 of these solutions (i.e. 5-6\% of the total) have core masses below 25 M$_\oplus$.

Before moving further, it should be emphasised that these solutions are obtained for the standard and enhanced opacity cases from \cite{LopFor14}, which refer to envelopes with solar metallicity or 50$\times$ solar metallicity, respectively. Envelope compositions intermediate between these two end-member cases, like that of Saturn \citep{atreya2018}, are not necessarily accurately fitted by these models. Further studies with more refined interior models are required before we can precisely assess the likelihood of Saturn-like or smaller cores. Notwithstanding this limitation, the relative frequency of solutions with Saturn-like or smaller cores in our Monte Carlo study does not allow to discard this possibility a priori.

While our pebble accretion simulations appear to point to core masses smaller than that estimated for Saturn \citep{mankovich2021}, those core masses are actually lower limits set only by the pebble isolation mass profile of the disk. Two processes, not included in the population synthesis study, can increase the final core mass of the planet and allow for Saturn-like interior structures of TOI-837\,$b$. First, the native protoplanetary disk of TOI-837\,$b$ could have been characterised by the presence of both pebbles and planetesimals at the time the giant planet formed. The accretion of planetesimals by TOI-837\,$b$ during its growth and migration would allow the core to grow beyond the pebble isolation mass and would enrich the envelope in heavy elements \citep{shibata2020,turrini2021}. 

Second, in a pebble-dominated disk the volatile elements are expected to sublimate from the drifting pebbles and enrich the disk gas in the orbital regions inward of the snowlines \citep{booth2019}. The accretion of such enriched gas would increase the budget of heavy elements of the giant planet and, if these heavy elements concentrate in the inner envelope, mimic the effects of a larger, diluted core like those suggested by interior studies of Jupiter \citep{wahl2017,stevenson2020} and Saturn \citep{mankovich2021}. 

While they can produce similar effects on the interior structure of the giant planet, these two processes have different compositional implications that can be used to further probe the formation history and interior structure of TOI-837\,$b$ through the atmospheric characterisation of its refractory-to-volatile ratio. Specifically, planetesimal accretion is more effective in enriching the giant planet in refractory elements, while pebble accretion is more effective in enriching it in volatile elements \citep{turrini2021,schneider2021,pacetti2022,crossfield2023}. Consequently, if the giant planet formed in a pebble-dominated disk, its atmosphere should show significant enrichment in C, O and N and limited or no enrichment in refractory elements. If the giant planets formed in a planetesimal-rich disk, due to its relatively high equilibrium temperature its atmospheres should show larger enrichment in, e.g., Na, K and S than in C, O, and N.

\subsection{Atmospheric evolution through photo-evaporation}
\label{sec:photoevap}

In order to study the atmospheric photo-evaporation over time, we followed a modelling approach initially proposed by \cite{Locci19}, which has been improved in several more recent works, and most recently described in detail in \citet{Mantovan24b}.

In brief, we evaluated the mass-loss rate of the planetary atmosphere by using the analytical approximation derived from the ATES hydrodynamic code \citep{Caldiroli+2021,Caldiroli+2022}, coupled with the planetary core-envelope models by \cite{Fortney2007} and \cite{LopFor14}, the MESA Stellar Tracks (MIST; \citealt{choi+2016}), the X-ray luminosity time evolution by \citet{Penz08a}, and the X-ray to EUV scaling by \citet{SF22}. 

To perform simulations of the past and future planetary evolution, we explored two core compositions, namely rock-iron (67\%, 33\%) and ice-rock (25\%, 75\%); the first is an Earth-like core, the second a Saturn-like core. For computing the planetary radius we  took into account two values of atmospheric metallicity, i.e.\ standard and enhanced  metallicity (see \citealt{LopFor14}). Therefore we simulated a total of 4 cases that we named rocki-std and rocki-hi for the standard and enhanced metallicity case, respectively, with an Earth-like core, and icer-std and icer-hi for the standard and enhanced metallicity case, respectively, with a Saturn-like core.

\begin{figure*}[h!]
\begin{tabular}{ccc} 
\hspace{-1cm}\includegraphics[width=8.cm]{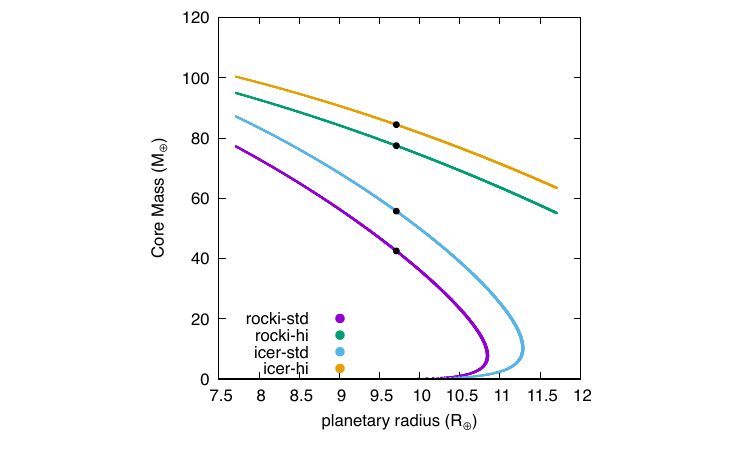} & \hspace{-2.5cm}\includegraphics[width=8.cm]{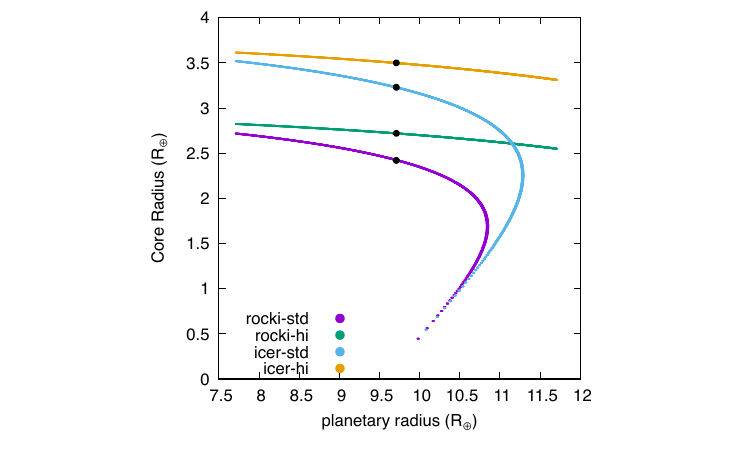} & \hspace{-2.5cm}\includegraphics[width=8.cm]{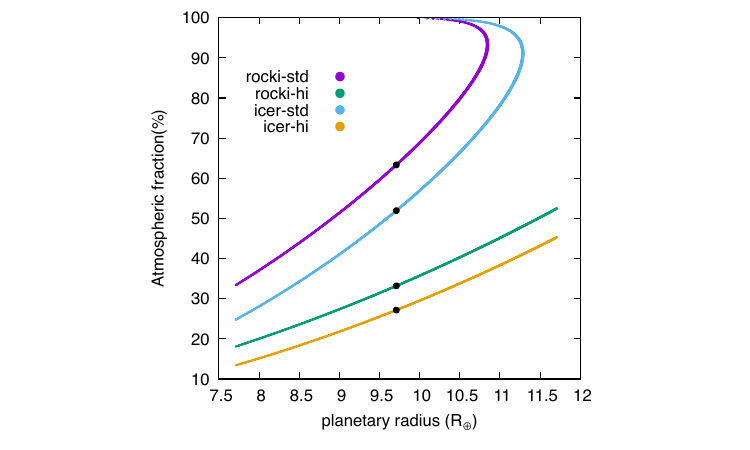} \\
\end{tabular}
\caption{Solutions of core-envelope models for planet TOI-837\,$b$ with total mass fixed at the measured value, considering both rock-iron core and ice-rock core scenarios with standard and enhanced metallicity. The left panel shows the core mass versus the planetary radius, the middle panel the core radius, and the right panel the atmospheric mass fraction.}
\label{fig:mrcore}
\end{figure*}

Figure \ref{fig:mrcore} shows the values of the core mass, core radius, and atmospheric mass fraction, at present age, as a function of the planetary radius for the various models. With the median values of mass and radius of TOI-837\,$b$ in Table \ref{tab:resultjointfit}, the system of equations that describe the planetary structure allows only one solution with a relatively large and massive core. Solutions with a smaller Saturn-like core can be found assuming a larger planetary radius within $1\sigma$ of the nominal value for an Earth-like composition, or slightly above $1\sigma$ for an ice-rock composition) and/or a smaller planetary mass (at least $3\sigma$ below our adopted value).
In Table \ref{table:par} we report the values we assumed in the following simulations for the mass and radius of the core, for each structure model. These values
remain constant in time, while the envelope radius  and atmospheric mass fraction may evolve in response to XUV irradiation. 

We determined the XUV flux at the planet, starting with the X-ray luminosity at current age. Since there is no direct measurement available, we estimated the X-ray luminosity from the Ca\,II H\&K chromospheric index (Sect.\ \ref{sec:chromindex}) following \citet{Mama+Hille2008}, and we obtained $\log L_{\rm x}$ in the range 3--$4 \times 10^{29}$\, erg\,s$^{-1}$. On the other hand, their relation with stellar age yields significantly larger values, near $\sim 1 \times 10^{30}$\, erg\,s$^{-1}$. We exclude such high values because of the lack of a X-ray detection of this source in the ROSAT All Sky Survey, and ultimately we adopted $L_{\rm x} = 5 \times 10^{29}$\, erg\,s$^{-1}$, as derived from the \citet{Pizz03} relation of X-ray luminosity to stellar rotation period.

\begin{figure*}[h!]
\begin{tabular}{ccc} 
\hspace{-1cm}\includegraphics[width=8.cm]{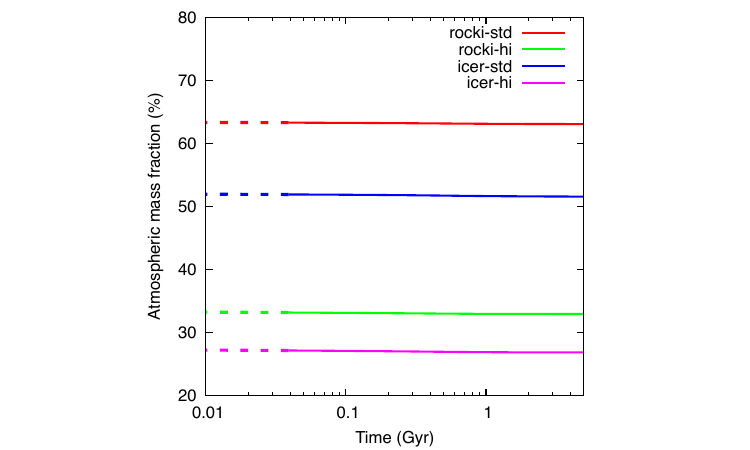} & \hspace{-2.5cm}\includegraphics[width=8.cm]{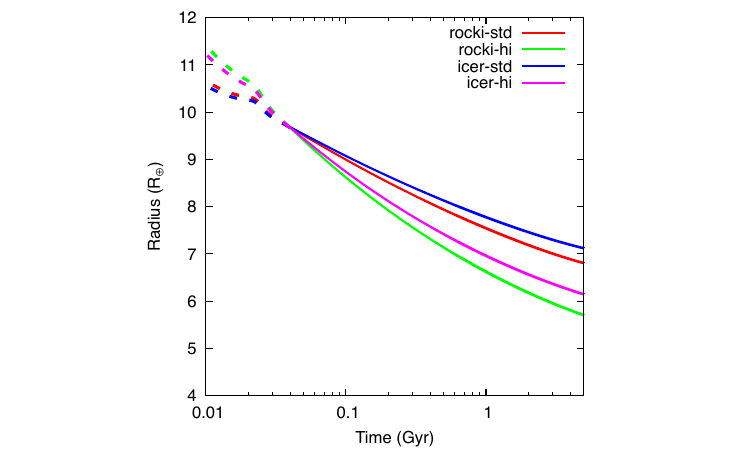} & \hspace{-2.5cm}\includegraphics[width=8.cm]{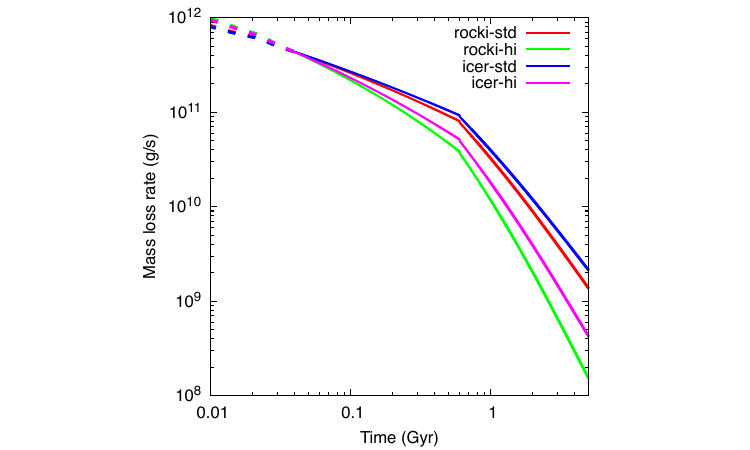} \\
\end{tabular}
\caption{Temporal evolution of mass fraction, radius, and mass-loss rate of TOI-837\,$b$. The left panel shows the evolution of atmospheric mass. fraction, the middle panel the evolution of the radius, and the right panel the evolution of the mass-loss rate. Solid lines represent the future evolution, whereas dashed lines the past one.}
\label{fig:evap}
\end{figure*}

Figure \ref{fig:evap} shows the evolution over time of the atmospheric mass fraction, planetary radius, and mass-loss rate. The values for these parameters at various ages are reported in Table \ref{table:par}. At present age we estimated a mass-loss rate of $\sim 5 \times 10^{11}$\,g/s, i.e.\ $\sim 2.5 \times 10^{-3}$\,\mearth/Myr.

We found that the planet is only slightly affected by hydrodynamic escape during the evolution, due to its large mass, which remains basically unchanged over time in every model we explored.
The mass-loss rate, being proportional to the cube of the radius (see \citealt{Caldiroli+2021,Caldiroli+2022}), is higher for models with larger radii, and of the order of $10^{12}$\,g/s at young ages (3--10\,Myr), but these values do not yield significant changes in the atmospheric fraction.
Consequently, the evolution of the radius is dominated by the natural gravitational contraction. According to these simulations, the planetary radius was 20--40\% larger at an age of 3\,Myr, and it will become 20--30\% smaller at an age of 1\,Gyr. We show in the mass-radius diagram of Fig. \ref{fig:mrdiagram} the expected positions at an age of 1\,Gyr for our predicted minimum and maximum values of the radius. If future follow-up observations will demonstrate that TOI-837\,$b$ has a structure compatible with a rock-iron core and a high opacity atmosphere (see Sect. \ref{sec:jwst}), this means that, according to our simulations, the planet will move to a region poorly populated by mature planets, making TOI-837\,$b$ an even more interesting object to characterise.

Furthermore, we performed an additional simulation assuming a radius of 10.64 \rearth\,(i.e. $1\sigma$ from the adopted value) in order to reproduce a Saturn-like core of about 20 \mearth\,with a rock-iron composition. Despite the larger radius and consequently higher mass-loss rate compared to previous simulations, also in this case we found that the evolutionary history of the atmosphere is scarcely affected by photo-evaporation due to the high planet's gravity.

\begin{table*}
    \scriptsize
    \caption{Results of the photo-evaporation modeling.}
    \label{table:par}     
    \centering              
    \begin{tabular}{c c|c c c c|c c c c|c c c c}       
	\hline    
    \noalign{\smallskip}
    \noalign{\smallskip}
    \textbf{Core Radius} &\textbf{ Core Mass} & \textbf{Mass }& \textbf{Radius} & \textbf{$f_{\rm atm}$} & \textbf{mass-loss rate} & \textbf{Mass} & \textbf{Radius} & \textbf{$f_{\rm atm}$} &\textbf{ mass-loss rate} & \textbf{Mass} & \textbf{Radius} & \textbf{$f_{\rm atm}$} &\textbf{ mass-loss rate} \\     
    (\rearth) & (\mearth) & ($M_\oplus$) & ($R_\oplus$) & (\%) & (g/s)  & ($M_\oplus$) & ($R_\oplus$) & (\%) & (g/s) & ($M_\oplus$) & ($R_\oplus$) & (\%) & (g/s) \\    
    \hline
     \noalign{\smallskip}                   
    \noalign{\smallskip}
    \multicolumn{2}{c}{Rock-iron core/Low opacity} & \multicolumn{4}{c}{current age} & 
    \multicolumn{4}{c}{at 3\,Myr}& \multicolumn{4}{c}{at 1\,Gyr} \\
     \noalign{\smallskip}
    \noalign{\smallskip}
     2.4 & 42.5 & 116.0 & 9.7 & 63.3 & $4.4 \times 10^{11}$ & 116.1 & 12.0 & 63.4 & $1.5 \times 10^{12}$ & 115.4 & 7.5 & 63.1&  $3.2 \times 10^{10}$ \\
    \hline
    \noalign{\smallskip}                   
    \noalign{\smallskip}
        \multicolumn{2}{c}{Rock-iron core/High opacity} & \multicolumn{4}{c}{current age} & \multicolumn{4}{c}{at 3\,Myr}& \multicolumn{4}{c}{at 1\,Gyr}\\
         \noalign{\smallskip}
    \noalign{\smallskip}
     2.7 & 77.5 & 116.0 & 9.7 & 33.2 & $4.4\times 10^{11}$ & 116.4 & 13.7 & 33.3 & $2.0 \times 10^{12}$ & 115.6 & 6.6 & 33.0 & $1.2\times 10^{10}$  \\
    \hline
    \noalign{\smallskip}
     \noalign{\smallskip}
    \multicolumn{2}{c}{Rock-ice core/Low opacity} & \multicolumn{4}{c}{current age} & \multicolumn{4}{c}{at 3\,Myr} & \multicolumn{4}{c}{at 1\,Gyr}\\
     \noalign{\smallskip}
    \noalign{\smallskip}
    3.2 & 55.7  & 116.0 & 9.7 & 51.9 & $4.4 \times 10^{11}$  & 116.1 & 11.7 & 52.0 & $1.4 \times 10^{12}$  & 115.4 & 7.8 & 51.7 & $4.0 \times 10^{10}$  \\  
    \hline  
     \noalign{\smallskip}
    \noalign{\smallskip}
    \multicolumn{2}{c}{Rock-ice core/High opacity} & \multicolumn{4}{c}{current age} & \multicolumn{4}{c}{at 3\,Myr}& \multicolumn{4}{c}{at 1\,Gyr}\\
     \noalign{\smallskip}
    \noalign{\smallskip}
     3.5 & 84.5  & 116.0 & 9.7 & 27.2 & $4.4 \times 10^{11}$  & 116.1 & 13.2 & 27.3 & $1.8 \times 10^{12}$  & 115.6 & 7.0 & 26.9 & $1.8 \times 10^{10}$  \\ 
    \hline   
    \end{tabular}
\end{table*}

\subsection{Atmospheric characterisation with \textit{JWST}} \label{sec:jwst}

In Fig. \ref{fig:mrdiagram} (second panel), we show the population of planets younger than 200 Myr and their transmission spectroscopy metric (TSM, \citealt{2018PASP..130k4401K}) values. TSM quantifies how much a transiting planet is amenable for atmospheric characterisation through transmission spectroscopy with the \textit{James Webb Space Telescope} (\textit{JWST}). Among all giant planets ($R_{\rm p} > 8~R_{\oplus}$, $M_{\rm p} > 0.1~M_{\rm J}$) orbiting stars younger than 200 Myr, TOI-837\,$b$ has the highest TSM value ($\sim$ 120), which is greater than the threshold (TSM=90) identified by \cite{2018PASP..130k4401K} for sub-Jovian planets. In principle, this makes TOI-837\,$b$ an ideal candidate for atmospheric characterisation by JWST. 

To quantify the prospects for atmospheric characterisation of TOI-837\,$b$ with JWST, we employed the open-source \textsc{Pyrat Bay} modeling framework \citep{CubillosBlecic2021mnrasPyratBay} to generate a variety of synthetic transmission and emission spectra. We fed the spectra into the \textsc{Pandeia} exposure time calculator \citep{PontoppidanEtal2016spiePandeia} to simulate transit and eclipse signals as observed by the JWST instruments and their S/N.

\begin{figure*}[h!] 
\includegraphics[width=\linewidth,clip]{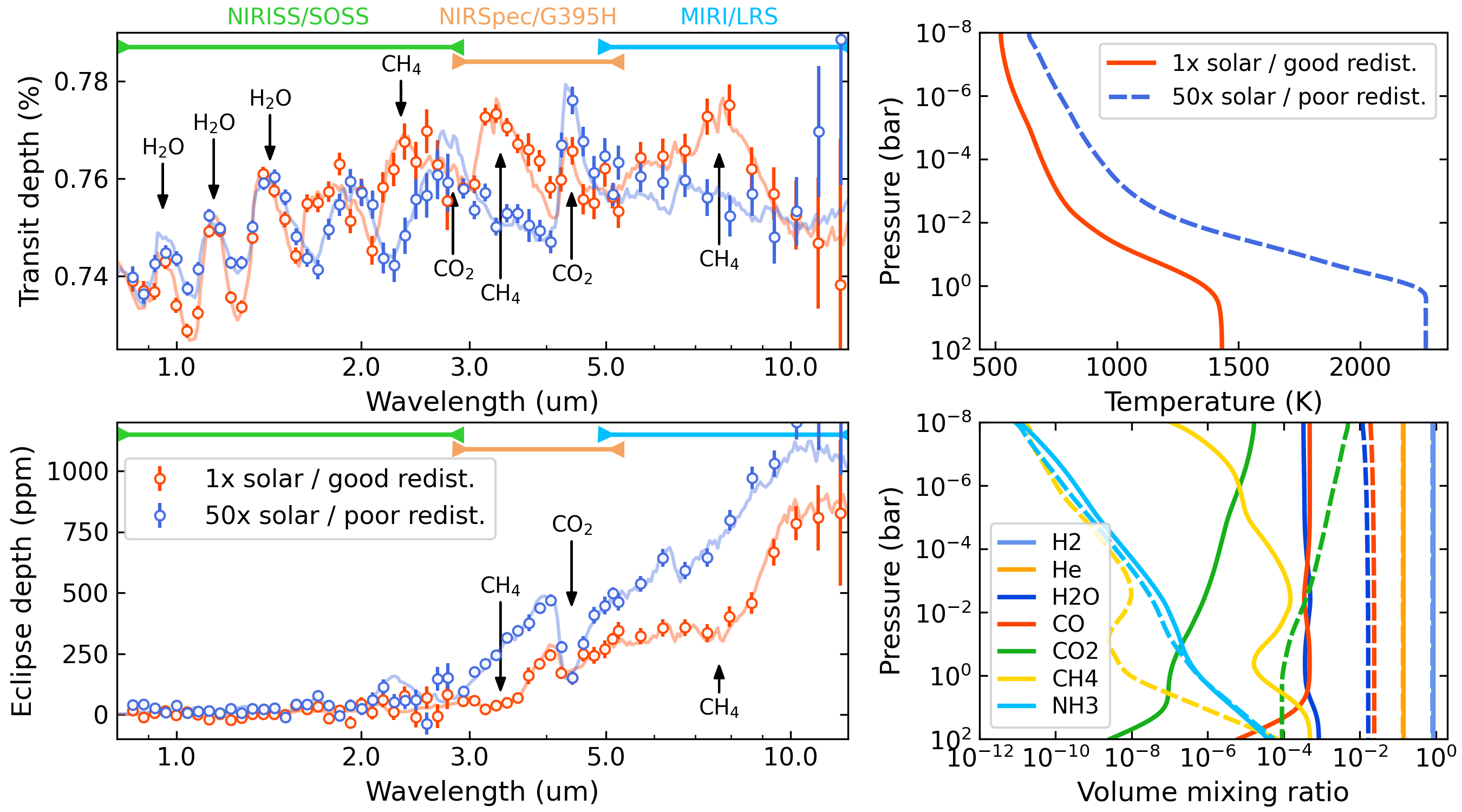}
\caption{JWST observation simulations of TOI-837\,$b$.  \textit{Left panels:} transit (top) and eclipse (bottom) simulations of JWST spectra for two scenarios: a solar-metallicity atmosphere with efficient day--night energy redistribution (red) and a 50$\times$ solar-metallicity atmosphere with poor energy redistribution (blue).  The markers with error bars show one realisation of an observation with the expected uncertainties (assuming one single observation per instrument). The coloured labels at the top denote the instrument and spectral range of the simulated observations. The solid curves denote the underlying `true' astrophysical models. The main molecular bands discussed in the text are indicated.  \textit{Top-right panel:} radiative-thermochemical-equilibrium temperature profiles which are the basis for the simulations and atmospheric modeling. \textit{Bottom-right panel:} Atmospheric abundances for the 1$\times$ (solid curves) and 50$\times$ solar metallicity model (dashed curves) for selected species (see legend).  }
\label{fig:jwst_simulations}
\end{figure*}

The \textsc{Pyrat Bay} package enables self-consistent atmospheric modeling (pressure, temperature, and composition) under an iterative radiative and thermochemical equilibrium calculation for a set of assumed physical conditions (Cubillos et al., in prep.). The equilibrium temperature of $\sim$1000 K makes TOI-837\,$b$ a particularly valuable target for atmospheric characterisation, since slight changes in the temperature or metallicity can lead to widely different compositions, each which clear and distinct observable outcomes. Therefore, to emphasise the extent to which the observable properties of TOI-837\,$b$ can change, we explored scenarios spanning the range of expected atmospheric metallicities and irradiation/dynamics conditions. The right panels of Fig. \ref{fig:jwst_simulations} show the temperature and composition profiles for two selected scenarios. The first scenario has a solar metallicity, zero Bond albedo, and a efficient day--night energy redistribution \citep[i.e., a heat-redistribution factor of $f=1/4$; see for example,][]{SpiegelBurrows2010apjModelSpectra}. The second scenario has an atmospheric metallicity of 50$\times$ solar, zero Bond albedo, and poor day--night energy redistribution ($f=2/3$).
As expected, the model with higher energy redistribution (1$\times$ solar model) has a cooler temperature profile (a difference of a few $\sim$100~K). The main consequence of this lower temperature is the proportionally larger \ch{CH4} abundance for the 1$\times$ solar model compared to the 50$\times$ solar model. On the other hand, the main consequence of a higher metallicity is the proportionally larger \ch{CO2} abundance. Both of these factors have a direct impact on the observable properties of the planet for either transit or eclipse observations.

For the JWST simulations, we initially considered all possible spectroscopic modes available: given the host star's spectral type and magnitude, all instruments can observe the system without reaching saturation.  We found the NIRISS/SOSS, NIRSpec/G395H, and MIRI/LRS settings as the optimal configurations that provide the better S/N and broader spectral coverage.
The left panels of Fig. \ref{fig:jwst_simulations} show the simulated transmission and emission observations for our two scenarios, considering a single transit/occultation and a single realisation per instrument, accounting for random noise added to the data. In transmission all three instruments can provide valuable abundance constraints, in particular: \ch{H2O} has a series of bands at the shorter wavelength end of the spectrum, being probed by NIRISS/SOSS; \ch{CO2} has its strongest absorption band at 4.4~{\microns}, being probed by NIRSpec/G395H; and \ch{CH4} has multiple bands that are readily detectable by all three instruments.
In emission, the lower planetary flux at the shorter wavelengths make NIRISS/SOSS the most challenging observation; however, both NIRSpec/G395H and MIRI/LRS can provide clear detection of atmospheric species like \ch{CO2} and \ch{CH4}.

Given the S/N of these simulations, JWST not only can detect the presence (or absence) of multiple molecular features on TOI-837\,$b$, but it can also constrain the abundance of specific species (along with the atmospheric thermal structure) by measuring the amplitude of their absorption bands. Finding muted spectral features would instead indicate the presence of clouds and hazes. Lastly, these observations also have the potential to probe the presence of refractory elements, which would constrain the formation and evolution of the system. For example, NIRISS/SOSS transmission observations can probe the signature of K at 0.77~{\microns} (not shown in Fig. \ref{fig:jwst_simulations}). NIRSpec and MIRI can probe the signature of \ch{SO2} at 4.0 and 7.5~{\microns}, respectively, which could become observable at sufficiently high abundances. In fact, in this way these observations could also become a probe for disequilibrium-chemistry processes, since processes like vertical transport and photochemistry are precisely a path to enhance the abundance of certain species including \ch{SO2} \citep[][]{TsaiEtal2023natWASP39bPhotochemistry}.

\section{Summary and prospects}
In this work we analysed and characterised the infant planetary system TOI-837 detected by TESS in the open cluster IC 2602, measuring the fundamental properties of the host star and of the close-in planet TOI-837\,$b$ through the modeling of photometric (TESS) and spectroscopic (HARPS) datasets. We found that TOI-837\,$b$ has a mass, radius, and bulk density which are similar to those of Saturn ($m_b$=116$^{+17}_{-18}$ \mearth; $r_b$=9.71$^{+0.93}_{-0.60}$ \rearth; $\rho_\oplus$= 0.68$^{+0.20}_{-0.18}$ \gcm). We did not find evidence for additional planetary companions in the system. Nonetheless, we identified a low-mass star which is likely gravitationally bound to TOI-837, and orbits at a separation of $\sim$330 au.   
The discovery of an infant Saturn-mass planet (stellar age $\sim$35 Myr) that closely orbits its parent star ($a_b \sim$0.08 au) induces fundamental questions about the formation and evolutionary histories of this planetary system. In our study, we first investigated possible formation scenarios and the predicted internal structure of TOI-837\,$b$ compatible with the measured mass and radius, then we reconstructed the evolution of the planet's atmosphere driven by photo-evaporation. In the following, we summarise our findings:    
\begin{itemize}
    \item Our formation and early migration models suggest both solutions with a relatively large and massive core, or a smaller Saturn-like core, depending on the opacity of the protoplanetary gas and on the growth rate of the core. 
    \item For the cases of standard opacity, we obtained synthetic planets with radius, mass and density within 1$\sigma$ of the best-fit (median) values in $\sim$30$\%$ of the possible interior structures, and 5-6$\%$ of the total synthetic planets have core masses below 25 \mearth.
    \item At its current age, because of the uncertainties on the planetary parameters and stellar age, the planet could have a structure with a massive core (40--80\,\mearth, depending on the actual core composition and opacity of the envelope), and a thick and heavy atmosphere (6--7\,\rearth, and 30--60\% the total mass), with a relatively low mass-loss rate ($2.5 \times 10^{-3}$\,\mearth/Myr). The possible solutions with a Saturn-like core and larger atmospheric mass fractions require a larger planetary radius (within $1\sigma$ from our nominal value in the case of a core with Earth-like composition), or a smaller mass.
    \item In any of the possible cases, our simulations of the evolutionary history indicate that the planet was little affected by photo-evaporation, even at early ages (3--10\,Myr) due to the relatively weak XUV irradiation and deep gravitational potential well. We conclude that, throughout its lifetime, TOI-837\,$b$ is expected to change slightly its position on the mass-radius diagram, unless follow-up observations will reveal that the actual planet's structure is compatible with a rocky-iron core and a high opacity atmosphere. In that case, we predict that TOI-837\,$b$ will move to a region of the diagram poorly populated by mature planets.
\end{itemize}

Based on our results, interesting questions arise that make TOI-837 a primary target for further observational follow-up. We advocate in particular for a more accurate and precise radius determination of TOI-837\,$b$ trough high-precision and high-angular resolution photometry (e.g. with the CHEOPS space telescope) to disentangle between the possible internal structures predicted by our population synthesis model. The same scientific case strongly motivates a follow-up of TOI-837 with JWST. As we demonstrated, TOI-837\,$b$ is a perfect target for transmission/emission spectroscopy with JWST, that will help to constrain composition and molecular abundances of the atmospheric envelope and, consequently, the actual internal structure of the planet. 

Being a short-period giant planet, TOI-837\,$b$ is a suitable target to measure its 3D obliquity with respect to the stellar rotation axis by detecting the Rossiter-McLaughlin (RM) effect \citep{1924ApJ....60...15R,1924ApJ....60...22M,2005ApJ...622.1118O}. With an expected amplitude of the RM signature $>$10 \ms, the target is amenable for a follow-up with high-resolution spectrographs such as ESPRESSO, which allowed for the detection of the RM effect in the case of DS\,Tuc\,A\,$b$ \citep{2021A&A...650A..66B}. So far, the RM effect has been detected for a handful of stars with ages $\le 100$\,Myr, and each of the obliquity measurements is consistent with zero (e.g. \citealt{2021ApJ...922L...1H,2022PASP..134h2001A}). It is important to verify this trend by measuring the obliquity of TOI-837\,$b$, to support the scenario that the close-in planet  underwent disk migration, as suggested by its orbital architecture. Alternatively, the gravitational pull exerted by the bound stellar companion to TOI-837 (discussed in Sect. \ref{sec:bin}) might have excited a high orbital obliquity of planet $b$.


\begin{acknowledgements}
We acknowledge the anonymous referee for insightful comments.
DPo acknowledges the support from the Istituto Nazionale di Oceanografia e Geofisica Sperimentale (OGS) and CINECA through the program ``HPC-TRES (High Performance Computing Training and Research for Earth Sciences)'' award number 2022-05 as well as the support of the  ASI-INAF agreement n 2021-5-HH.1-2022. AMa acknowledges partial support from the PRIN-INAF
2019 (project HOT-ATMOS), and the ASI-INAF agreement n 2021.5-HH.1-2022.  PCu is funded by the Austrian Science Fund (FWF) Erwin Schroedinger Fellowship, program J4595-N. GMa acknowledges support from CHEOPS ASI-INAF agreement n. 2019-29-HH.0. This work has been supported by the PRIN-INAF 2019 ``Planetary systems at young ages (PLATEA)''.
\end{acknowledgements}

%
%
%
\bibliographystyle{aa}
\bibliography{toi837}
\begin{appendix}
\section{Data}
\longtab[1]{
\begin{longtable}{ccc}
\caption{HARPS radial velocities.}          
\label{table:dataHS} \\     
\hline\hline      
\noalign{\smallskip}
Time & RV & $\sigma_{\rm RV}$  \\    
(BJD) & (\ms) & (\ms) \\ 
\noalign{\smallskip}
\hline\hline
\endfirsthead
\caption{Continued.}\\
\hline\hline      
\noalign{\smallskip}
Time & RV & $\sigma_{\rm RV}$ \\    
(BJD) & (\ms) & (\ms) \\ 
\noalign{\smallskip}
\hline\hline
\endhead
\hline
\endfoot
\hline  
\endlastfoot
\noalign{\smallskip}
  2459859.874407&  -8.00&  36.25\\
 2459861.882975&  54.06&  46.85\\
 2459863.880785&  -2.07&  27.64\\
 2459874.868108& -39.52&  11.14\\
 2459875.863933& -44.49&  15.77\\
 2459876.862513& 186.06&  12.58\\
 2459877.872957&  -3.95&  35.77\\
 2459878.867058&  15.40&  18.60\\
 2459879.864431& 225.50&  11.48\\
 2459881.828297&   4.36&  11.12\\
 2459882.851749& 204.35&  11.11\\
 2459883.852314&  88.23&  29.24\\
 2459884.806778&  98.00&  26.53\\
 2459885.813024& 244.01&  17.73\\
 2459887.858511&  51.08&  10.59\\
 2459888.841509& 246.24&  12.29\\
 2459891.848433& 147.09&   9.59\\
 2459893.827244&  27.61&  20.20\\
 2459898.852755&  12.75&   8.54\\
 2459899.859217& -29.81&   9.04\\
 2459900.865879&  89.32&   8.92\\
 2459902.826645&  17.07&   6.31\\
 2459903.834429& 275.11&  10.29\\
 2459904.834460& -21.73&   8.73\\
 2459905.803170&  11.36&   8.55\\
 2459906.841983& 215.46&  10.57\\
 2459907.818849& -76.82&  12.23\\
 2459908.779705& -12.53&   9.24\\
 2459909.837371& 178.01&  10.16\\
 2459927.727831& 139.60&  21.67\\
 2459927.748623&  77.93&  26.31\\
 2459927.812161& 184.12&  15.58\\
 2459927.832121& 142.72&  18.66\\
 2459929.763470&  11.01&  10.24\\
 2459929.841509& -17.01&   7.34\\
 2459930.726162& 191.34&  13.25\\
 2459930.836015& 164.22&  19.02\\
 2459931.774526& -79.08&   9.02\\
 2459931.852518& -64.97&   8.17\\
 2459933.846959& 129.50&  12.10\\
 2459934.736238& -97.46&   8.55\\
 2459934.853508& -68.67&   7.96\\
 2459935.753849&   0.59&   6.63\\
 2459935.768997&   9.36&  10.87\\
 2459936.686875& 227.08&   8.55\\
 2459936.839195& 203.82&  10.07\\
 2459937.698063& -48.04&   7.97\\
 2459937.811657& -15.47&   7.35\\
 2459938.651974& 100.49&  10.64\\
 2459938.845089&  21.58&   9.39\\
 2459939.768104& 215.66&  11.23\\
 2459939.841857& 181.50&  13.10\\
 2459941.829967& -33.88&   7.34\\
 2459942.728895& 210.87&   9.40\\
 2459942.840736& 160.36&   7.83\\
 2459943.699822& -64.13&   9.97\\
 2459943.833009& -34.92&   9.75\\
 2459946.735117&   5.60&   6.75\\
 2459946.837326&  52.94&   9.33\\
 2459947.745071&   2.03&   7.95\\
 2459947.838803& -29.04&   9.33\\
 2459955.770815& -21.04&   8.93\\
 2459956.796814& -33.54&   8.57\\
 2459957.720902& 312.16&   9.09\\
 2459957.856480& 276.67&   9.05\\
 2459958.786049& -63.24&   8.54\\
 2459958.863078& -20.30&   9.50\\
 2459959.694040&  -5.07&  12.53\\
 2459959.837175& -46.54&  10.86\\
 2459960.747339& 330.03&   9.65\\
\hline                                   
\end{longtable}
}
\begin{small}
\longtab[2]{
\begin{longtable}{ccccccccc}
\caption{Time series of the activity diagnostics considered in this work.}          
\label{table:actdiagnostics} \\     
\hline\hline      
\noalign{\smallskip}
Time & S$_{\rm MW}$ & $\sigma_{\rm S_{\rm MW}}$ & H-alpha & $\sigma_{\rm H-alpha}$ & DLW & $\sigma_{\rm DLW}$ & CRX & $\sigma_{\rm CRX}$ \\    
(BJD) &  &  &  &  & (m$^2$/s$^2$) & (m$^2$/s$^2$) & (\ms) & (\ms)\\ 
\noalign{\smallskip}
\hline\hline
\endfirsthead
\caption{Continued.}\\
\hline\hline      
\noalign{\smallskip}
Time & S$_{\rm MW}$ & $\sigma_{\rm S_{\rm MW}}$ & H-alpha & $\sigma_{\rm H-alpha}$ & DLW & $\sigma_{\rm DLW}$ & CRX & $\sigma_{\rm CRX}$ \\    
(BJD) &  &  &  &  & (m$^2$/s$^2$) & (m$^2$/s$^2$) & (\ms) & (\ms)\\ 
\noalign{\smallskip}
\hline\hline
\endhead
\hline
\endfoot
\hline  
\endlastfoot
\noalign{\smallskip}
2459859.874407&0.3394&0.0032&0.2328&0.0010&  935.83&  112.70&-1211.87&  253.27\\
2459861.882975&0.3067&0.0035&0.2346&0.0011&  790.54&  154.03&-1628.57&  314.39\\
2459863.880785&0.3135&0.0028&0.2293&0.0009&  803.66&  121.03& -846.43&  196.45\\
2459874.868108&0.3603&0.0025&0.2421&0.0008& -491.49&   80.99&  147.22&   90.09\\
2459875.863933&0.3155&0.0025&0.2336&0.0008& -547.47&  101.81&  354.21&  121.68\\
2459876.862513&0.3276&0.0026&0.2289&0.0008& -632.62&   92.10&  107.47&  103.38\\
2459877.872957&0.3322&0.0038&0.2421&0.0012&-1499.41&  199.08& 1434.70&  229.23\\
2459878.867058&0.3305&0.0024&0.2326&0.0008& -413.23&   87.80&  754.45&  118.52\\
2459879.864431&0.3432&0.0020&0.2307&0.0006& -181.35&   48.16&  183.09&   93.03\\
2459881.828297&0.3421&0.0017&0.2319&0.0005&   88.82&   35.40&  368.71&   80.52\\
2459882.851749&0.3280&0.0018&0.2311&0.0006&  -22.76&   59.29&   79.28&   93.35\\
2459883.852314&0.3366&0.0032&0.2421&0.0010& -751.55&  142.44& 1284.86&  184.79\\
2459884.806778&0.3520&0.0028&0.2388&0.0009& -198.67&  103.48&  931.00&  190.19\\
2459885.813024&0.3307&0.0024&0.2336&0.0007&  -57.97&   90.66&  306.30&  145.04\\
2459887.858511&0.3423&0.0016&0.2315&0.0005&  383.51&   50.59&  343.95&   77.73\\
2459888.841509&0.3380&0.0019&0.2315&0.0006&  278.01&   76.50&   -0.84&  105.18\\
2459891.848433&0.3371&0.0016&0.2327&0.0006&  522.70&   67.78&   -6.04&   82.93\\
2459893.827244&0.3461&0.0025&0.2373&0.0008&  849.66&   82.99&  548.62&  155.38\\
2459898.852755&0.3643&0.0015&0.2416&0.0005&  581.44&   55.47&  128.33&   71.52\\
2459899.859217&0.3482&0.0015&0.2334&0.0005&  841.02&   50.05&   88.47&   76.47\\
2459900.865879&0.3475&0.0016&0.2335&0.0006&  886.86&   55.16&  -78.09&   75.16\\
2459902.826645&0.3469&0.0014&0.2334&0.0005&  775.51&   44.35&  106.45&   51.54\\
2459903.834429&0.3356&0.0016&0.2302&0.0005&  821.96&   50.93& -433.00&   66.36\\
2459904.834460&0.3655&0.0017&0.2383&0.0006&  799.05&   53.21&  228.20&   67.84\\
2459905.803170&0.3566&0.0018&0.2362&0.0006& 1025.68&   63.43&  -94.43&   71.35\\
2459906.841983&0.3340&0.0016&0.2308&0.0006&  880.26&   60.67& -401.62&   73.35\\
2459907.818849&0.3476&0.0021&0.2368&0.0007&  982.23&   75.74&  142.32&  102.03\\
2459908.779705&0.3596&0.0018&0.2354&0.0006& 1035.85&   56.32&  112.14&   77.34\\
2459909.837371&0.3446&0.0014&0.2320&0.0005&  715.03&   56.23& -372.94&   72.63\\
2459927.727831&0.3467&0.0027&0.2364&0.0008& -445.66&   64.09& -796.95&  145.59\\
2459927.748623&0.3290&0.0032&0.2417&0.0010&-1114.68&  102.32&-1063.61&  165.98\\
2459927.812161&0.3524&0.0021&0.2358&0.0007&  -75.26&   41.12& -563.07&  105.93\\
2459927.832121&0.3465&0.0023&0.2387&0.0007& -306.93&   52.75& -766.11&  116.56\\
2459929.763470&0.3502&0.0017&0.2335&0.0006&  -36.78&   34.61& -145.26&   83.50\\
2459929.841509&0.3402&0.0014&0.2351&0.0005&  262.70&   27.13&   24.46&   61.65\\
2459930.726162&0.3437&0.0020&0.2366&0.0006& -288.81&   36.86& -503.39&   87.59\\
2459930.836015&0.3462&0.0024&0.2357&0.0007& -560.87&   53.72& -736.76&  122.55\\
2459931.774526&0.3415&0.0018&0.2316&0.0006& -282.23&   37.68&  -43.97&   74.66\\
2459931.852518&0.3494&0.0018&0.2347&0.0006& -268.04&   38.50&  -38.99&   67.68\\
2459933.846959&0.3476&0.0020&0.2366&0.0007& -442.38&   40.53& -360.19&   87.85\\
2459934.736238&0.3421&0.0018&0.2327&0.0006& -343.41&   42.14&  202.60&   65.32\\
2459934.853508&0.3410&0.0018&0.2356&0.0006& -607.99&   34.55& -105.75&   64.30\\
2459935.753849&0.3351&0.0016&0.2331&0.0005&  -17.51&   34.63&    4.46&   55.18\\
2459935.768997&0.3420&0.0023&0.2303&0.0008&    2.63&   36.02& -134.96&   89.21\\
2459936.686875&0.3566&0.0017&0.2376&0.0005& -228.35&   28.03& -418.07&   45.57\\
2459936.839195&0.3515&0.0017&0.2388&0.0006& -307.31&   37.09& -384.36&   66.83\\
2459937.698063&0.3381&0.0017&0.2334&0.0005& -339.33&   30.30&  182.67&   61.46\\
2459937.811657&0.3379&0.0017&0.2322&0.0006& -354.07&   36.80&   92.06&   59.54\\
2459938.651974&0.3316&0.0028&0.2364&0.0008&-1058.21&   85.03& -164.06&   83.88\\
2459938.845089&0.3488&0.0024&0.2353&0.0008& -792.60&   71.48&   12.91&   76.60\\
2459939.768104&0.3322&0.0026&0.2378&0.0008&-1036.71&   73.21& -372.57&   76.76\\
2459939.841857&0.3223&0.0026&0.2405&0.0009&-1179.04&   81.70& -371.34&   94.62\\
2459941.829967&0.3442&0.0018&0.2328&0.0006& -254.64&   44.03&  218.00&   53.63\\
2459942.728895&0.3474&0.0018&0.2378&0.0006& -404.39&   36.33& -290.64&   67.43\\
2459942.840736&0.3538&0.0017&0.2405&0.0005& -298.29&   34.15& -228.83&   57.97\\
2459943.699822&0.3450&0.0020&0.2362&0.0006& -569.41&   40.60&  251.45&   75.10\\
2459943.833009&0.3349&0.0020&0.2349&0.0007& -587.76&   51.92&  239.73&   73.75\\
2459946.735117&0.3470&0.0018&0.2360&0.0006& -457.41&   37.43&  138.09&   52.58\\
2459946.837326&0.3432&0.0021&0.2362&0.0007& -677.55&   50.24&  210.94&   71.06\\
2459947.745071&0.3346&0.0016&0.2307&0.0005&  -77.29&   36.02&  262.13&   56.61\\
2459947.838803&0.3359&0.0018&0.2310&0.0006& -265.95&   35.73&  375.25&   59.31\\
2459955.770815&0.3354&0.0016&0.2350&0.0006& -285.03&   39.07&  227.63&   68.17\\
2459956.796814&0.3269&0.0015&0.2339&0.0005&   57.16&   29.47&  307.34&   59.50\\
2459957.720902&0.3289&0.0020&0.2382&0.0007& -584.40&   54.05& -209.12&   69.25\\
2459957.856480&0.3414&0.0022&0.2393&0.0007& -650.95&   45.24&  -67.74&   73.33\\
2459958.786049&0.3386&0.0016&0.2370&0.0005& -154.14&   32.20&  293.76&   59.81\\
2459958.863078&0.3326&0.0018&0.2348&0.0005& -250.68&   29.80&  271.65&   70.12\\
2459959.694040&0.3239&0.0019&0.2352&0.0006& -223.65&   41.18&  474.67&   82.15\\
2459959.837175&0.3210&0.0020&0.2335&0.0006& -261.79&   42.55&  337.25&   77.72\\
2459960.747339&0.3283&0.0020&0.2388&0.0007& -594.17&   50.07& -286.29&   70.75\\
\hline                                   
\end{longtable}
}
\end{small}
\section{Additional plots} 
\begin{figure*}[]
    \centering
    \begin{subfigure}{\linewidth}
    \centering
    \includegraphics[width=\textwidth]{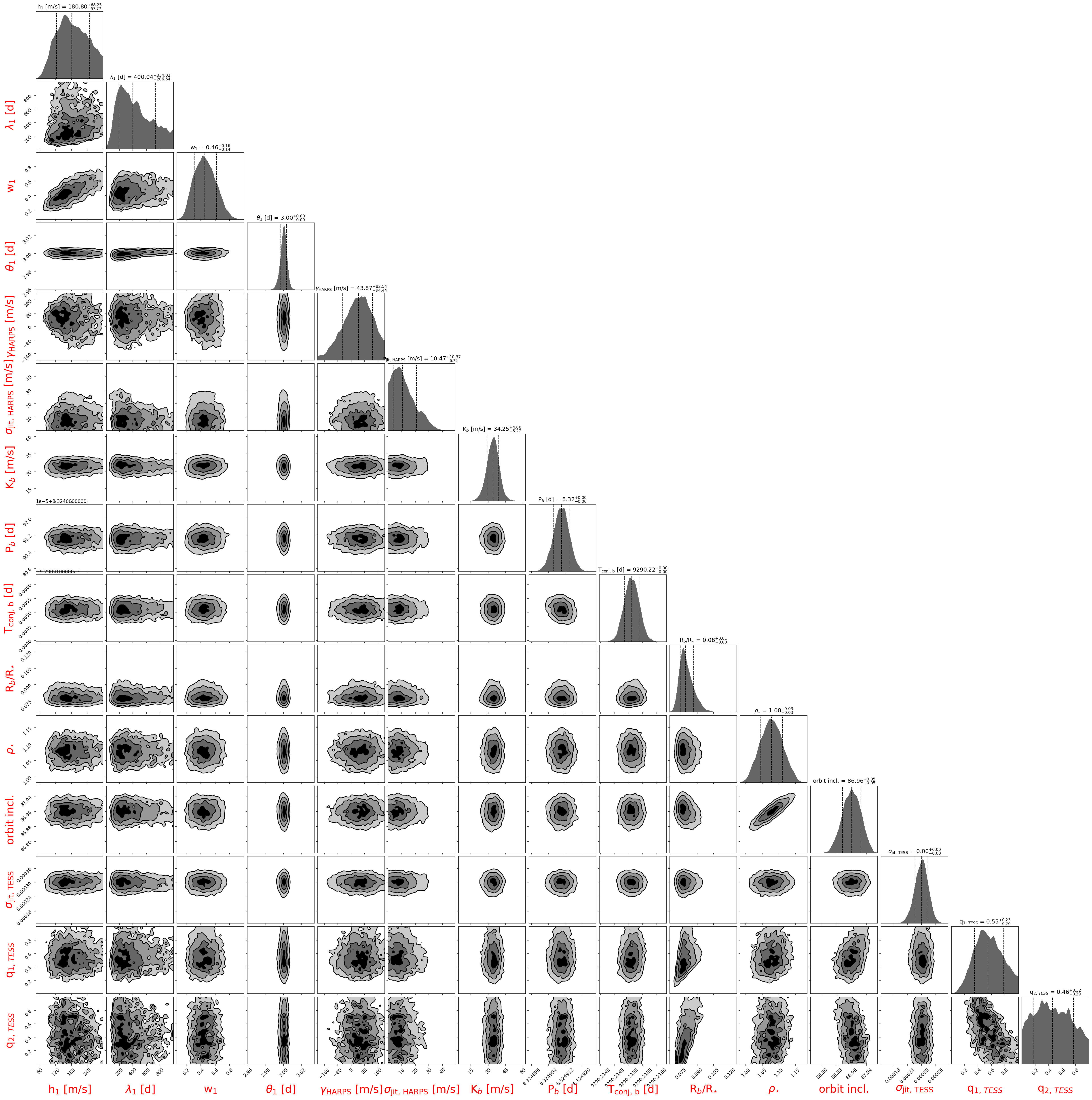}\\
    \caption{}
    \end{subfigure}
 \end{figure*}
 
\clearpage  
\begin{figure*}    
 \addtocounter{figure}{-1}
    \begin{subfigure}{\textwidth}
    \addtocounter{subfigure}{1} 
    \centering
    \includegraphics[width=\textwidth]{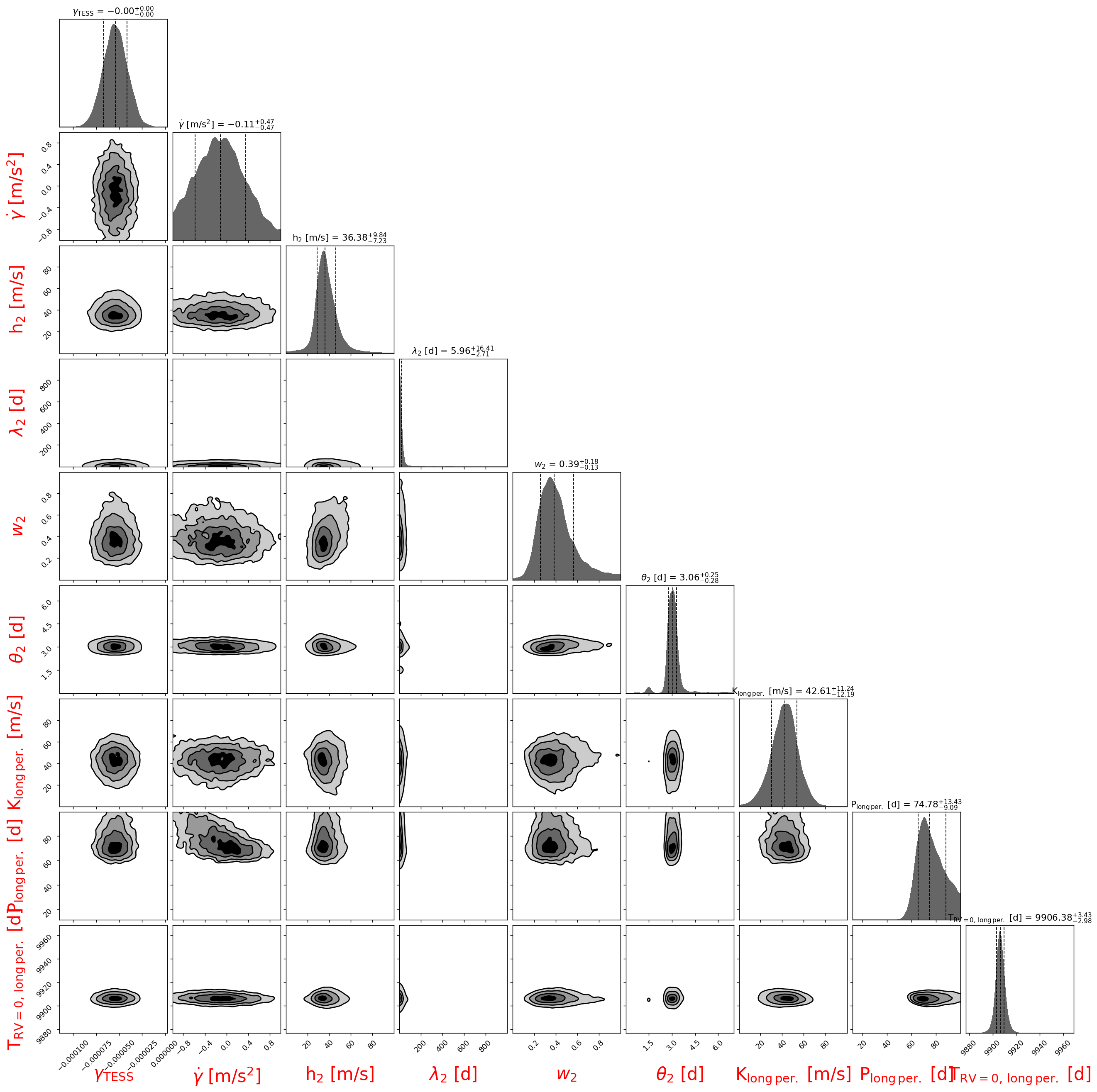}
    \caption{}
    \end{subfigure}
      \caption{Corner plot showing the posterior distributions of the free (hyper-)parameters for our adopted joint RV+light curve model discussed in Sect. \ref{sec:rvphotoanalysis} (two GP quasi-periodic signals plus two sinusoids). Instead of a single corner plot, two subplots (a) and (b) are shown for a better readability, preserving the visibility of correlations for a few parameters.}
          \label{fig:cornerplot}
\end{figure*}

\begin{figure*}[h!]
    \centering
    \includegraphics[width=0.7\linewidth]{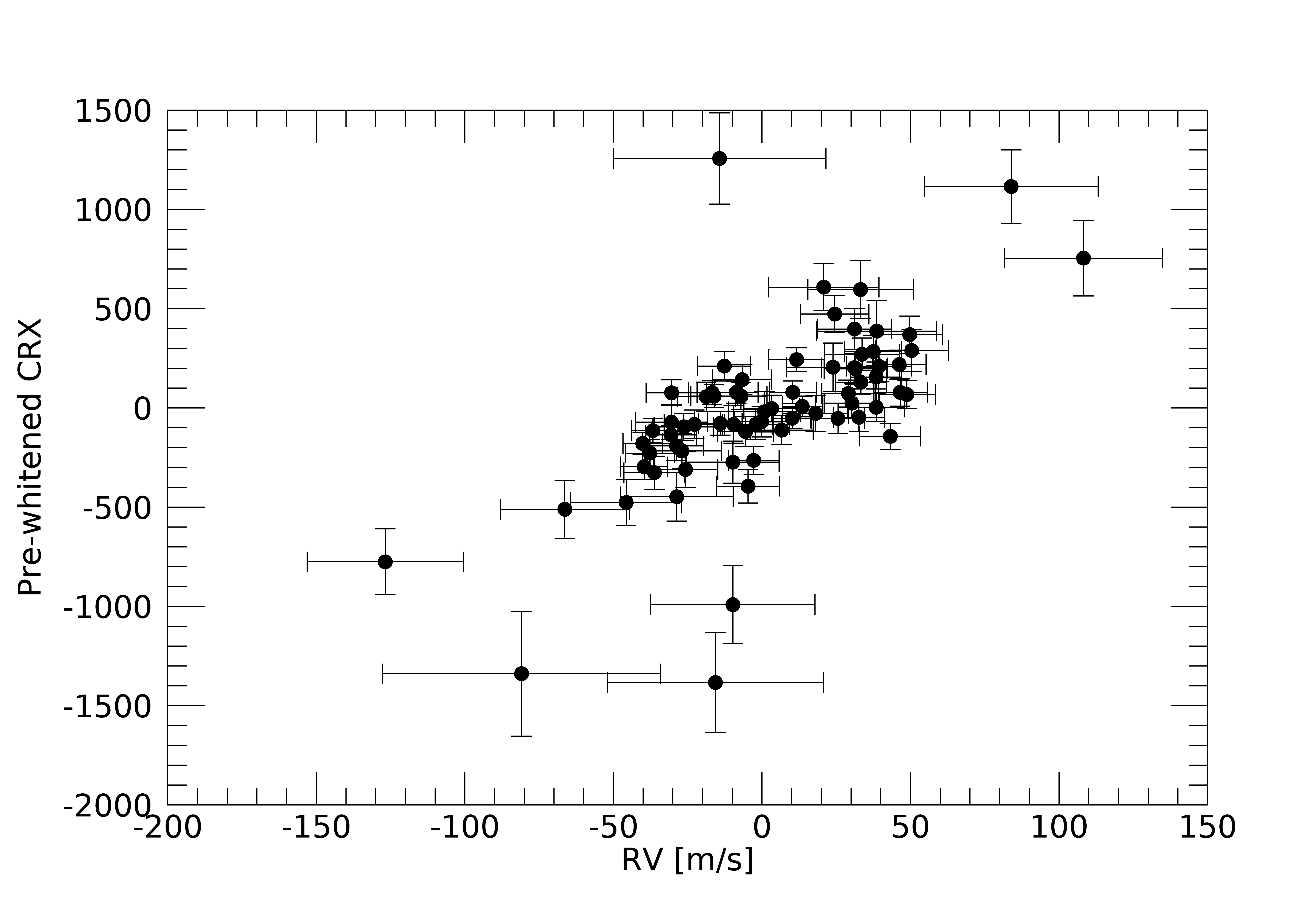}
    \caption{Pre-whitened CRX activity index plotted versus the RVs of the long-period sinusoid included in our adopeted best-fit RV model. The correlation coefficient is $\rho_{\rm Pearson}$=+0.71, indicating that the RV signal is very likely due to stellar activity.}
    \label{fig:rvcrxcorr}
\end{figure*}
\end{appendix}
\end{document}